\documentclass[11pt]{article}
\usepackage[colorlinks=true,citecolor=blue,linkcolor=blue]{hyperref}
\usepackage{multirow}% for tables
\usepackage[left=2cm,right=2cm,top=2cm,bottom=2cm]{geometry}
\usepackage{cite}
\usepackage{psfrag}
\usepackage[demo]{graphicx}
\usepackage{graphicx}
\usepackage{graphics}
\usepackage{epsfig}
\usepackage{amsmath,amsfonts,amssymb}
\usepackage{pstcol,pst-fill,pst-grad}
\usepackage{pstricks,pst-fill,pst-grad}
\usepackage{euscript}
\usepackage{pstricks}
\usepackage{wrapfig}
\usepackage{slashed}
\usepackage[toc,page]{appendix}
\usepackage{here}
\date{}
\begin{document}
	%\begin{titlepage}
	\title{\vspace{-3cm}
		\hfill\parbox{4cm}{\normalsize \emph{}}\\
		\vspace{1cm}
		{ New phenomena in laser-assisted leptonic decays of the negatively charged boson $W^{-}$}}
	\vspace{2cm}
	
	\author{ S Mouslih$^{2,1}$, M Jakha$^{1}$, I Dahiri$^{1}$, S Taj$^1$, B Manaut$^{1,}$\thanks{Corresponding author, E-mail: b.manaut@usms.ma}  and E. Siher$^{2}$ \\
		{\it {\small$^1$ Sultan Moulay Slimane University, Polydisciplinary Faculty,}}\\
		{\it {\small Research Team in Theoretical Physics and Materials (RTTPM), Beni Mellal, 23000, Morocco.}}\\			
	{\it {\small$^2$Faculty of Sciences and Techniques, 
		Laboratory of Materials Physics (LMP),
		Beni Mellal, 23000, Morocco.}}		
	}
	\maketitle \setcounter{page}{1}



\date{\today}

\begin{abstract}
The majority of studies and experiments performed at electron-positron colliders over the last two decades have focused on studying $W$ and $Z$ weak force-carrying bosons and accurately measuring all their properties, not only because they play an important role in establishing Standard Model theory and providing an accurate test of its predictions of particle interactions, but also because they are a unique tool for probing manifestations of the new physics beyond the standard model. Therefore, it would be particularly important to discuss some of the new phenomena and changes that can arise in these bosons when their decay occurs under an external electromagnetic field. In a recent paper, we investigated the laser effect on the final products of $Z$ boson decay and found that laser had an unprecedented effect on branching ratios. In this work and within the standard Glashow-Weinberg-Salam model of electroweak interactions, we study theoretically the leptonic decay of the $W^{-}$-boson $(W^{-}\rightarrow \ell^{-}\bar{\nu}_{\ell})$ in the presence of a circularly polarized electromagnetic field and we examine the laser effect, in terms of its field strength and frequency, on the leptonic decay rate and the phenomenon of multiphoton processes. The calculations are carried out using the exact relativistic wave functions of charged particles in an electromagnetic field. It was found that the laser significantly contributed to reducing the probability of $W^{-}$-boson decay. We show that the laser-assisted decay rate is equal to the laser-free one only when the famous Kroll-Watson sum rule is fulfilled. The notable effect of the laser on the leptonic decay rate was reasonably interpreted by the well-known quantum Zeno effect or by the opening of channels other than leptonic ones to decay. This work will pave the way for an upcoming one to study the hadronic decay of the $W^{-}$-boson and then explore the laser effect on its lifetime and branching ratios.
\end{abstract}

\maketitle
\section{Introduction}
The tremendous and rapid progress that laser technology has made since its invention in the $1960$s has enabled a new field of theoretical and experimental studies to explore the interactions of electromagnetic (EM) fields with matter at high intensities \cite{salamin}. This important development of laser technology has played an important role in the production of high-power laser systems with a maximum intensity of $10^{22}\text{W/cm}^{2}$ \cite{bahk,yanovski} and researchers currently expect that higher intensities can be achieved in the near future \cite{baifei,tajima,bulanov,pukhov}. Today, these high-power laser systems have become indispensable equipment and are used in laboratories all over the world, thanks to which studies of the basic processes of quantum electrodynamics (QED) \cite{qed1,qed2,hartin,ritus1,ritus2,ehlotzky} and electroweak theory \cite{ritus3,becker} in the presence of a strong laser field and realistic in experiments have continued to attract significantly the attention of physicists. Through all this, scientists are seeking to know and understand how elementary particles behave and how their properties change when inserted in an EM field. The theory describing the fundamental interactions of these particles is called the Standard Model (SM) and it is a gauge theory, which describes strong, weak and electromagnetic interactions by exchanging the corresponding spin-$1$ gauge fields: eight massless gluons and one massless photon, respectively, for the strong and electromagnetic interactions, and three massive bosons, $W^{\pm}$ and $Z^{0}$, for the weak interaction. It is one of the most successful achievements of modern physics in that it provides a very elegant theoretical framework capable of describing known experimental facts in particle physics with great precision. The discovery of the $W$ \cite{ua1w,ua2w} and $Z$ \cite{ua1z,ua2z} bosons in $1983$ by the UA1 and UA2 collaborations at the CERN's $p\bar{p}$ collider provided direct confirmation of the unification of the weak and electromagnetic interactions in a common framework that describes them together. This common framework is the so-called electroweak theory predicted by physicists Sheldon Lee Glashow \cite{glashow}, Steven Weinberg \cite{weinberg} and Abdus Salam \cite{salam} in the late $1960$s, in which they explain that the electromagnetic and weak forces, which have long been considered as separate entities, are in fact manifestations of the same fundamental interaction. In recognition of their role in the discovery of the $W$ and $Z$ particles, the CERN physicist Carlo Rubbia \cite{rubbia} and engineer Simon van der Meer \cite{vander} were awarded the Nobel Prize in physics in $1984$. Detectable $W$ particles can be produced in proton-antiproton collision experiments by processes such as $u\bar{d}$ or $\bar{u}d\rightarrow W$ followed by subsequent leptonic $(W\rightarrow \ell \nu)$ or hadronic decay $(W\rightarrow q \bar{q}')$, where $\ell=e,~\mu,~\tau$ and $q$ or $q'$ represent one of the quarks $u$, $d$, $c$, $s$, or $b$ except $t$ since top quark is heavier than the $W$ boson. Experiments at the Large Electron-Positron Collider (LEP) and the Stanford Linear Collider (SLC) refine the measurements of the $W$-boson properties (mass, total decay rate and cross sections of their production). At the same time, the properties of $W$ bosons are studied in the Tevatron proton-antiproton accelerator. All these experiments have given more precise measurements of the $W$-boson mass and its total decay rate, which are now known to be $M_{W}=80.379\pm0.012~\text{GeV}$ and $\Gamma_{W}=2.085\pm 0.042~\text{GeV}$ \cite{pdg2020}. The accuracy obtained in these experiments makes it possible to test the predictions of the SM at the level of radiative corrections. This work has, as an objective, to study the leptonic decay of the negatively charged boson $W^{-}$ in the presence of a circularly polarized EM field. As mentioned above, the $W^{-}$-boson can decay into a lepton and antineutrino $W^{-}\rightarrow \ell^{-}\bar{\nu}_{\ell}$ (leptonic decay) or into a pair of quarks $W^{-}\rightarrow q \bar{q}'$ (hadronic decay). Here, we will only consider the first channel (leptonic channel), where we will try to realize the effect of the EM field strength and frequency on the decay rate. In $2004$, the author Kurilin studied the leptonic decays of the $W^{-}$-boson in a strong crossed EM Field, where he calculated the expression of the decay rate and studied its changes in terms of the external-field-strength parameter $\varkappa=eM_{W}^{-3}\sqrt{-(F_{\mu\nu}q^{\nu})^{2}}$, where $F_{\mu\nu}$ is the EM tensor \cite{kurilin2004}. Kurilin's approach is completely different from the approach followed here. In his study, Kurilin restricts his consideration to the case of a so-called crossed field, which is a superposition of constant electric and magnetic fields whose strength vectors are equal and orthogonal. The wave functions of the charged particles in this case have a simple expression in terms of the EM tonsor $F_{\mu\nu}$, and this makes the related mathematical calculations less complicated. Some of the transformations in our calculations have been simplified by the introduction of the ordinary Bessel functions, whose order is interpreted as the number of exchanged photons, while Kurilin calculations have been performed with the help of special mathematical functions generically termed Airy functions \cite{abramowitz}. Despite this difference in the method and the nature of the EM field, we have compared our results obtained here with the results obtained in his research and give some explanations for both results obtained. There is also an earlier paper who did the same study in brief giving the expression of the decay rate in the presence of a circularly polarized EM field \cite{obukhov1}. An overview of the interaction of these bosons with the EM field and how to write the expression for the wave functions describing them can be found in refs. \cite{kurilin1999,obukhov2}. Although $W$ bosons are best known for their role in radioactive decay, their decay may provide the opportunity to discover supersymmetric particles and probe new physics beyond the SM \cite{bsm1,bsm2}. The present work is part of a series of studies we have conducted to know and understand the effect of the EM field on experimentally measurable quantities during electroweak decay processes (decay rate, lifetime, branching ratio). So far, we have calculated both the pion \cite{mouslih} and $Z$-boson \cite{jakha} decays in the presence of an EM field in an attempt to contribute to enriching the debate on this topic. In this same context, we try to study the decay of the $W^{-}$-boson in two stages. First, we will study its leptonic decay and thus the effect of the EM field on the leptonic decay rate. Secondly, in a forthcoming research, we will study the hadronic decay and collect the results obtained with the present results in order to be able to study the effect of the laser on both the lifetime and the branching ratio for each channel as two important points that must be treated. It is therefore this work that will pave the way for the next one, and by that time we will have studied this decay and have sufficiently covered all its aspects. The rest of the paper is arranged as follows. In section \ref{sec:theory}, we will try to formulate, in detail, the theoretical expression of the decay rate, then in section \ref{sec:results}, we will present the results obtained and highlight the extent of the laser influence on the decay rate. The last section \ref{sec:conclusion} is devoted to summarizing the conclusions that we have reached. We only mention here that we used, during our calculations, known natural units $c=\hbar=1$. 
\section{Theoretical formulation of the decay rate in a laser field}
\label{sec:theory}
\begin{figure}[hbtp]
\centering
\includegraphics[scale=0.5]{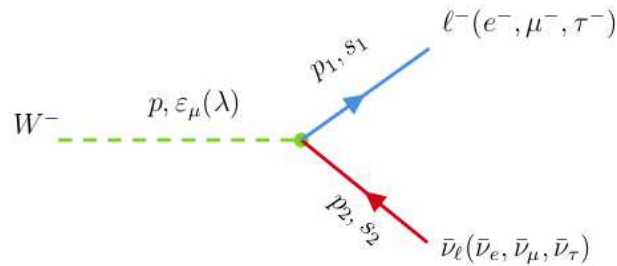}
\caption{Lowest order Feynman diagram of the leptonic $W^{-}$-boson decay.}\label{diagram}
\end{figure}

We begin by considering the process of electroweak decay of a boson vector $W^{-}$, with four-momentum $p$ and mass $M_{W}$, into a lepton $\ell^{-}$ and corresponding antineutrino $\bar{\nu}_{\ell}$ in the field of a plane EM wave
\begin{equation}\label{process}
 W^{-}(p)\longrightarrow \ell^{-}(p_{1})+\bar{\nu}_{\ell}(p_{2}),~~~(\ell=e,\mu,\tau)
\end{equation}
where the arguments are our labels for the four-momenta. The corresponding Feynman diagram is shown in figure \ref{diagram}.
The laser field, which is assumed to be circularly polarized monochromatic, is described by the following classical four-potential
\begin{align}\label{potential}
A^{\mu}(\phi)=a^{\mu}_{1}\cos(\phi)+a^{\mu}_{2}\sin(\phi),
\end{align}
depending on a single variable, $\phi=(k.x)$, the phase of the laser field. $k=(\omega,\textbf{k})$ is the wave four-vector $(k^{2}=0)$ and $\omega$ is the laser frequency. The  four-amplitudes $a^{\mu}_{1}=|\textbf{a}|(0,1,0,0)$ and $a^{\mu}_{2}=|\textbf{a}|(0,0,1,0)$ are equal in magnitude and orthogonal, which implies $(a_{1}.a_{2})=0$ and $a_{1}^{2}=a_{2}^{2}=a^{2}=-|\textbf{a}|^{2}=-(\mathcal{E}_{0}/\omega)^{2}$ where $\mathcal{E}_{0}$ is the amplitude of the electric field. We suppose that the four-potential satisfies the Lorentz gauge condition, $k_{\mu}A^{\mu}=0$, which means $(k.a_{1})=(k.a_{2})=0$, indicating that the wave vector $\textbf{k}$ is chosen to be along the $z$-axis. \\
The lowest-order scattering S-matrix element for the laser-assisted leptonic $W^{-}$ decay reads \cite{greiner}
\begin{equation}\label{smatrix}
S_{fi}(W^{-}\rightarrow \ell^{-}\bar{\nu}_{\ell})=\dfrac{i\textsl{g}}{2\sqrt{2}}\int d^{4}x\overline{\psi}_{\ell}(x)\gamma^{\mu}(1-\gamma_{5})\psi_{\bar{\nu}_{\ell}}(x)W^{-}_{\mu}(x),
\end{equation}
where $\textsl{g}$ is the electroweak coupling constant. $\psi_{\ell}(x)$ is the wave function of the relativistic lepton $\ell^{-}$ in an EM field given by the relativistic Dirac-Volkov functions normalized to the volume V \cite{volkov}
\begin{equation}\label{lwavefunction}
\psi_{\ell}(x)=\bigg[1+\dfrac{e\slashed{k}\slashed{A}}{2(k.p_{1})}\bigg]\frac{u(p_{1},s_{1})}{\sqrt{2Q_{1}V}}\times e^{iS(q_{1},x)},
\end{equation}
where regarding the sign one should note that $e=-|e|$ is the charge of the electron, and
\begin{equation}
S(q_{1},x)=-q_{1}.x-\dfrac{e(a_{1}.p_{1})}{k.p_{1}}\sin(\phi)+\dfrac{e(a_{2}.p_{1})}{k.p_{1}}\cos(\phi).
\end{equation}
$u(p_{1},s_{1})$ represents the Dirac bispinors for the free lepton with momentum $p_{1}$ and spin $s_{1}$ satisfying $\sum_{s_{1}}u(p_{1},s_{1})\overline{u}(p_{1},s_{1})=\slashed{p}_{1}+m_{\ell}$
where $m_{\ell}$ is the rest mass of the lepton.
The 4-vector $q_{1}=(Q_{1},\textbf{q}_{1})$ is the quasi-momentum  that the lepton acquires in the presence of the EM field
\begin{equation}
q_{1}=p_{1}-\frac{e^{2}a^{2}}{2(k.p_{1})}k,~~~q_{1}^{2}=m^{*2}_{\ell}=m_{\ell}^{2}-e^{2}a^{2},
\end{equation}
where $m^{*}_{\ell}$ is an effective mass of the lepton inside the EM field.\\
For the laser-dressed $W^{-}$ (1-spin particle), its wave function can be written in the form \cite{kurilin1999}
\begin{equation}\label{wwavefunction}
W^{-}_{\mu}(x)=\bigg[\textsl{g}_{\mu\nu}-\dfrac{e}{(k.p)}\big(k_{\mu}A_{\nu}-k_{\nu}A_{\mu}\big)-\frac{e^{2}}{2(k.p)^{2}}A^{2}k_{\mu}k_{\nu}\bigg]\frac{\varepsilon_{\mu}(p,\lambda)}{\sqrt{2p_{0}V}}\times e^{iS(q,x)},
\end{equation}
where $\textsl{g}_{\mu\nu}=\text{diag}(1,-1,-1,-1)$ is the metric tensor of Minkowski space, $\varepsilon_{\mu}(p,\lambda)$ is the $W^{-}$-boson polarization four-vector such that the summation over all three directions of polarization $\lambda$ yields $\sum_{\lambda=1}^{3}\varepsilon_{\mu}(p,\lambda)\varepsilon^{*}_{\nu}(p,\lambda)=-\textsl{g}_{\mu\nu}+p_{\mu}p_{\nu}/M_{W}^{2}$, and
\begin{equation}
S(q,x)=-q.x+\dfrac{e(a_{1}.p)}{k.p}\sin(\phi)-\dfrac{e(a_{2}.p)}{k.p}\cos(\phi),
\end{equation}
with the 4-vector $q=p-\big[e^{2}a^{2}/2(k.p)\big]k$ $(q^{2}=M^{*2}_{W})$ and $M^{*}_{W}=\sqrt{M_{W}^{2}-e^{2}a^{2}}$ are, respectively, the quasi-momentum and the effective mass that the boson $W^{-}$ acquires inside the EM field.
The outgoing antineutrino $\bar{\nu}_{\ell}$ is treated as massless particle with four-momentum $p_{2}$ and spin $s_{2}$. According to the Feynman rules, it is represented by an incoming wave function with negative four-momentum as follows \cite{greiner}:
\begin{align}\label{neutrinowave}
\psi_{\bar{\nu}_{\ell}}(x)=\dfrac{v(p_{2},s_{2})}{\sqrt{2E_{2}V}}e^{ip_{2}.x},
\end{align}
where $E_{2}=p_{2}^{0}=|\textbf{p}_{2}|$ and $v(p_{2},s_{2})$ is the Dirac spinor satisfying $\sum_{s_{2}}v(p_{2},s_{2})\overline{v}(p_{2},s_{2})=\slashed{p}_{2}$.
Substituting the lepton, antineutrino and $W^{-}$-boson wave functions (\ref{lwavefunction}), (\ref{wwavefunction}) and (\ref{neutrinowave}) into expression (\ref{smatrix}) for the S-matrix element, we obtain
\begin{equation}\label{smatrix1}
\begin{split}
S_{fi}(W^{-}\rightarrow \ell^{-}\bar{\nu}_{\ell})=&\dfrac{i\textsl{g}}{2\sqrt{2}\sqrt{8E_{2}Q_{1}p_{0}V^{3}}}\int d^{4}x\overline{u}(p_{1},s_{1})\Big\lbrace\Big[1+C(p_{1})\slashed{A}\slashed{k}\Big]\gamma^{\mu}(1-\gamma_{5})\\&\times\Big[\textsl{g}_{\mu\nu}-C(p)\big(k_{\mu}A_{\nu}-k_{\nu}A_{\mu}\big)-\frac{C(p)^{2}}{2}a^{2}k_{\mu}k_{\nu}\Big]\Big\rbrace v(p_{2},s_{2})\\&\times\varepsilon_{\mu}(p,\lambda) e^{ip_{2}.x}e^{i(S(q,x)-S(q_{1},x))},
\end{split}
\end{equation}
where $C(p_{1})=e/(2(k.p_{1}))$ and $C(p)=e/(k.p)$. Now, let us transform the exponential term $e^{i(S(q,x)-S(q_{1},x))}$ by introducing the following parameter
\begin{equation}\label{argument}
z=\sqrt{\alpha_{1}^{2}+\alpha_{2}^{2}}\quad\text{with}\quad\alpha_{1}=-e\bigg(\dfrac{a_{1}.p}{k.p}+\dfrac{a_{1}.p_{1}}{k.p_{1}}\bigg)\,;\,\alpha_{2}=-e\bigg(\dfrac{a_{2}.p}{k.p}+\dfrac{a_{2}.p_{1}}{k.p_{1}}\bigg),
\end{equation}
this yields
\begin{equation}
e^{i(S(q,x)-S(q_{1},x))}=e^{i(q_{1}-q).x} e^{-iz\sin(\phi-\phi_{0})},
\end{equation}
with $\phi_{0}=\arctan(\alpha_{2}/\alpha_{1})$. Therefore, the S-matrix element becomes
\begin{equation}\label{smatrix2}
\begin{split}
S_{fi}(W^{-}\rightarrow \ell^{-}\bar{\nu}_{\ell})=&\dfrac{i\textsl{g}}{2\sqrt{2}\sqrt{8E_{2}Q_{1}p_{0}V^{3}}}\int d^{4}x\overline{u}(p_{1},s_{1})\big[C_{0}+ C_{1}\cos(\phi)+C_{2}\sin(\phi)\big]\\&\times v(p_{2},s_{2})\varepsilon_{\mu}(p,\lambda) e^{i(p_{2}+q_{1}-q).x}e^{-iz\sin(\phi-\phi_{0})},
\end{split}
\end{equation}
where the three quantities $C_{0}$, $C_{1}$, and $C_{2}$ are expressed as follows:
\begin{equation}
\begin{split}
C_{0}=&\gamma_{\nu}(1-\gamma_{5})-\frac{C(p)^{2}}{2}a^{2}k_{\mu}k_{\nu}\gamma^{\mu}(1-\gamma_{5}),\\
C_{1}=&C(p_{1})\slashed{a}_{1}\slashed{k}\gamma_{\nu}(1-\gamma_{5})-C(p)\gamma^{\mu}(1-\gamma_{5})\big(k_{\mu}a_{1\nu}-k_{\nu}a_{1\mu}\big)-\frac{C(p)^{2}}{2}C(p_{1})a^{2}k_{\mu}k_{\nu}\\&\times\slashed{a}_{1}\slashed{k}\gamma^{\mu}(1-\gamma_{5}),\\
C_{2}=&C(p_{1})\slashed{a}_{2}\slashed{k}\gamma_{\nu}(1-\gamma_{5})-C(p)\gamma^{\mu}(1-\gamma_{5})\big(k_{\mu}a_{2\nu}-k_{\nu}a_{2\mu}\big)-\frac{C(p)^{2}}{2}C(p_{1})a^{2}k_{\mu}k_{\nu}\\&\times\slashed{a}_{2}\slashed{k}\gamma^{\mu}(1-\gamma_{5}).
\end{split}
\end{equation}
The linear combination of the three different quantities in equation~(\ref{smatrix2}) can be transformed by the well-known identities involving ordinary Bessel functions  $J_{s}(z)$ \cite{landau}:
\begin{align}\label{transformation}
\begin{bmatrix}
1\\
\cos(\phi)\\
\sin(\phi)
\end{bmatrix}\times e^{-iz\sin(\phi-\phi_{0})}=\sum_{s=-\infty}^{+\infty}\begin{bmatrix}
B_{s}(z)\\
B_{1s}(z)\\
B_{2s}(z)
\end{bmatrix}e^{-is\phi},
\end{align}
where
\begin{align}
\begin{split}
\begin{bmatrix}
B_{s}(z)\\
B_{1s}(z)\\
B_{2s}(z) \end{bmatrix}=\begin{bmatrix}J_{s}(z)e^{is\phi_{0}}\\
\big(J_{s+1}(z)e^{i(s+1)\phi_{0}}+J_{s-1}(z)e^{i(s-1)\phi_{0}}\big)/2\\
\big(J_{s+1}(z)e^{i(s+1)\phi_{0}}-J_{s-1}(z)e^{i(s-1)\phi_{0}}\big)/2i
 \end{bmatrix},
\end{split}
\end{align}
where $z$ is the argument of the Bessel functions defined in equation~(\ref{argument}) and $s$, their order,  can be interpreted as the number of exchanged photons. Using these transformations in equation~(\ref{smatrix2}) and  integrating over $d^{4}x$, the matrix element $S_{fi}$ becomes
\begin{equation}\label{smatrix3}
S_{fi}(W^{-}\rightarrow \ell^{-}\bar{\nu}_{\ell})=\dfrac{i\textsl{g}}{2\sqrt{2}\sqrt{8E_{2}Q_{1}p_{0}V^{3}}}\sum_{s=-\infty}^{\infty}\mathcal{M}^{s}_{fi}(2\pi)^{4}\delta^{4}(p_{2}+q_{1}-q-sk),
\end{equation}
where the quantity $\mathcal{M}^{s}_{fi}$ is defined by
\begin{align}
\mathcal{M}^{s}_{fi}=\bar{u}(p_{1},s_{1})\Lambda_{s}v(p_{2},s_{2})\varepsilon_{\mu}(p,\lambda),
\end{align}
where
\begin{align}
\Lambda_{s}=C_{0}B_{s}(z)+ C_{1}B_{1s}(z)+C_{2}B_{2s}(z).
\end{align}
Here, we call the quantity $\mathcal{M}^{s}_{fi}$ "the spinorial part" since it is the only part of the matrix element which depends on spin.
Multiplying the squared S-matrix element by the density of final states, summing over spins of leptons and antineutrinos, averaging over the polarization of the incoming boson, and finally dividing by the time $T$ yields the decay rate
\begin{align}\label{summed}
\Gamma(W^{-}\rightarrow \ell^{-}\bar{\nu}_{\ell})=\sum_{s=-\infty}^{+\infty}\Gamma^{s}(W^{-}\rightarrow \ell^{-}\bar{\nu}_{\ell}),
\end{align}
where $\Gamma^{s}$, called the photon-number-resolved decay rate, is defined by
\begin{align}
\Gamma^{s}(W^{-}\rightarrow \ell^{-}\bar{\nu}_{\ell})=\dfrac{\textsl{g}^{2}}{64p_{0}}\int\dfrac{d^{3}q_{1}}{(2\pi)^{3}Q_{1}}\int\dfrac{d^{3}p_{2}}{(2\pi)^{3}E_{2}}(2\pi)^{4}\delta^{4}(p_{2}+q_{1}-q-sk)|\overline{\mathcal{M}^{s}_{fi}}|^{2},
\end{align}
where we have used $[(2\pi)^{4}\delta^{4}(p_{2}+q_{1}-q-sk)]^{2}= V T (2\pi)^{4}\delta^{4}(p_{2}+q_{1}-q-sk)$, and
\begin{align}
|\overline{\mathcal{M}^{s}_{fi}}|^{2}=\frac{1}{3}\sum_{\lambda}\sum_{s_{1},s_{2}}|\mathcal{M}^{s}_{fi}|^{2}=\frac{1}{3}\sum_{\lambda}\sum_{s_{1},s_{2}}|\bar{u}(p_{1},s_{1})\Lambda_{s}v(p_{2},s_{2})\varepsilon_{\mu}(p,\lambda)|^{2}.
\end{align} 
Performing the integration over $d^{3}p_{2}$ and using $\delta^{4}(p_{2}+q_{1}-q-sk)=\delta^{3}(\textbf{p}_{2}+\textbf{q}_{1}-\textbf{q}-s\textbf{k})\delta(E_{2}+Q_{1}-Q-s\omega)$, the photon-number-resolved decay rate $\Gamma^{s}$ becomes
\begin{align}
\Gamma^{s}(W^{-}\rightarrow \ell^{-}\bar{\nu}_{\ell})=\dfrac{\textsl{g}^{2}}{64(2\pi)^{2}p_{0}}\int\dfrac{d^{3}q_{1}}{Q_{1}E_{2}}\delta(E_{2}+Q_{1}-Q-s\omega)|\overline{\mathcal{M}^{s}_{fi}}|^{2},
\end{align}
with $\textbf{p}_{2}+\textbf{q}_{1}-\textbf{q}-s\textbf{k}=0$. We choose the $W^{-}$-boson rest frame in which $Q=q_{0}=M^{*}_{W}$ and $\textbf{q}=0$, then $\textbf{p}_{2}=s\textbf{k}-\textbf{q}_{1}$. Hence, using $ d^{3}q_{1}=|\textbf{q}_{1}|^{2}d|\textbf{q}_{1}|d\Omega_{\ell}$, we obtain
\begin{align}
\begin{split}
\Gamma^{s}(W^{-}\rightarrow \ell^{-}\bar{\nu}_{\ell})=&\dfrac{\textsl{g}^{2}}{64(2\pi)^{2}p_{0}}\int\dfrac{|\textbf{q}_{1}|^{2}d|\textbf{q}_{1}|d\Omega_{\ell}}{Q_{1}E_{2}}\delta\Big(\sqrt{(s\omega)^{2}+|\textbf{q}_{1}|^{2}-2s\omega|\textbf{q}_{1}|\cos(\theta)}\\&+\sqrt{|\textbf{q}_{1}|^{2}+m_{\ell}^{*2}}-M^{*}_{W}-s\omega\Big)|\overline{\mathcal{M}^{s}_{fi}}|^{2}.
\end{split}
\end{align}
The remaining integral over $d|\textbf{q}_{1}|$ can be solved by using the familiar formula \cite{greiner}:
\begin{align}\label{familiarformula}
\int dxf(x)\delta(g(x))=\dfrac{f(x)}{|g'(x)|}\bigg|_{g(x)=0}.
\end{align}
Thus we get
\begin{align}\label{ws0}
\Gamma^{s}(W^{-}\rightarrow \ell^{-}\bar{\nu}_{\ell})=&\dfrac{\textsl{g}^{2}}{64(2\pi)^{2}p_{0}}\int\dfrac{|\textbf{q}_{1}|^{2}|d\Omega_{\ell}}{Q_{1}E_{2}|g'(|\textbf{q}_{1}|)|}|\overline{\mathcal{M}^{s}_{fi}}|^{2}.
\end{align}
where $\textsl{g}^{2}=8G_{F}M_{W}^{2}/\sqrt{2}$, with $G_{F}=(1.166~37\pm0.000~02)\times10^{-11}~\text{MeV}^{-2}$ is the Fermi coupling constant, and
\begin{align}
g'(|\textbf{q}_{1}|)=\dfrac{|\textbf{q}_{1}|-s\omega\cos(\theta)}{\sqrt{(s\omega)^{2}+|\textbf{q}_{1}|^{2}-2s\omega|\textbf{q}_{1}|\cos(\theta)}}+\dfrac{|\textbf{q}_{1}|}{\sqrt{|\textbf{q}_{1}|^{2}+m_{\ell}^{*2}}}.
\end{align}
Using the fact that $d\Omega_{\ell}=\sin(\theta)d\theta d\varphi$ and $\int d\varphi=2\pi$, the equation (\ref{ws0}) can be written as follows:
\begin{align}\label{ws}
\Gamma^{s}(W^{-}\rightarrow \ell^{-}\bar{\nu}_{\ell})=&\dfrac{\textsl{g}^{2}}{128\pi p_{0}}\int\dfrac{|\textbf{q}_{1}|^{2}|\sin(\theta)d\theta}{Q_{1}E_{2}|g'(|\textbf{q}_{1}|)|}|\overline{\mathcal{M}^{s}_{fi}}|^{2},
\end{align}
where the integral over $d\theta$ is performed with the
help of numerical integration.\\
The term  $|\overline{\mathcal{M}^{s}_{fi}}|^{2}$ can be calculated by converting the sums over the spins into traces as follows:
\begin{align}\label{trace}
|\overline{\mathcal{M}^{s}_{fi}}|^{2}=\frac{1}{3}\bigg(-\textsl{g}_{\mu\nu}+\frac{p_{\mu}p_{\nu}}{M_{W}^{2}}\bigg)\text{Tr}\big[(\slashed{p}_{1}+m_{\ell})\Lambda_{s}\slashed{p}_{2}\overline{\Lambda}_{s}\big],
\end{align}
where
\begin{align}
\begin{split}
\overline{\Lambda}_{s}&=\gamma^{0}\Lambda_{s}^{\dagger}\gamma^{0},\\
&=\overline{C}_{0}B^{*}_{s}(z)+ \overline{C}_{1}B^{*}_{1s}(z)+\overline{C}_{2}B^{*}_{2s}(z),
\end{split}
\end{align}
and
\begin{equation}
\begin{split}
\overline{C}_{0}=&\gamma^{0}C_{0}^{\dagger}\gamma^{0}=\gamma_{\mu}(1-\gamma_{5})-\frac{C(p)^{2}}{2}a^{2}k_{\mu}k_{\nu}\gamma^{\nu}(1-\gamma_{5}),\\
\overline{C}_{1}=&\gamma^{0}C_{1}^{\dagger}\gamma^{0}=C(p_{1})\gamma_{\mu}(1-\gamma_{5})\slashed{k}\slashed{a}_{1}-C(p)\gamma^{\nu}(1-\gamma_{5})\big(k_{\mu}a_{1\nu}-k_{\nu}a_{1\mu}\big)\\&-\frac{C(p)^{2}}{2}C(p_{1})a^{2}k_{\mu}k_{\nu}\gamma^{\nu}(1-\gamma_{5})\slashed{k}\slashed{a}_{1},\\
\overline{C}_{2}=&\gamma^{0}C_{2}^{\dagger}\gamma^{0}=C(p_{1})\gamma_{\mu}(1-\gamma_{5})\slashed{k}\slashed{a}_{2}-C(p)\gamma^{\nu}(1-\gamma_{5})\big(k_{\mu}a_{2\nu}-k_{\nu}a_{2\mu}\big)\\&-\frac{C(p)^{2}}{2}C(p_{1})a^{2}k_{\mu}k_{\nu}\gamma^{\nu}(1-\gamma_{5})\slashed{k}\slashed{a}_{2}.
\end{split}
\end{equation}
The trace calculation is performed with the help of FEYNCALC \cite{feyncalc1,feyncalc2,feyncalc3}. The detailed and explicit expression obtained for the averaged squared "spinorial part" $|\overline{\mathcal{M}^{s}_{fi}}|^{2}$ is given by
\begin{equation}\label{result}
\begin{split}
|\overline{\mathcal{M}^{s}_{fi}}|^{2}=\dfrac{1}{3}\bigg[&AJ_{s}^{2}(z)+BJ_{s+1}^{2}(z)+CJ_{s-1}^{2}(z)+DJ_{s}(z)J_{s+1}(z)+EJ_{s}(z)J_{s-1}(z)\\&+FJ_{s+1}(z)J_{s-1}(z)\bigg],
\end{split}
\end{equation}
where the six coefficients $A$, $B$, $C$, $D$, $E$ and $F$ are explicitly expressed by
\begin{equation}
\begin{split}
A=&\dfrac{4}{(k.p)^2 M_{W}^2}\big[a^4 e^4 (k.p_{1}) (k.p_{2}) + 
   2 a^2 e^2 (2 (k.p_{1}) (k.p_{2}) M_{W}^2 + (k.p)^2 (p_{1}.p_{2}) \\
 &- (k.p) (k.p_{2}) (p.p_{1}) - (k.p) (k.p_{1}) (p.p_{2})) + 2 (k.p)^2 (M_{W}^2 (p_{1}.p_{2})\\
 & + 2 (p.p_{1}) (p.p_{2})) \big],
\end{split}
\end{equation}
\begin{equation}
\begin{split}
 B=&\dfrac{-e^2}{2 (k.p) (k.p_{1})^3 M_{W}^2} \big[ 4 (k.p) (k.p_{1})^2 (-(a_{1}.p_{1}) (a_{1}.p_{2}) (k.p) + a^2 ((k.p) (p_{1}.p_{2}) \\
 &+ (k.p_{2}) (M_{W}^2 - (p.p_{1})) + 2 (k.p) (p.p_{2}) + (k.p_{1}) (p.p_{2})))-a^2 (k.p_{1}) \\ 
 &\times(a^2 e^2 ((k.p) (k.p_{2}) p_{0} - (k.p_{1}) (k.p_{2}) p_{1}^{0} - 2 (k.p)^2 E_{2}) + 2(k.p) ((k.p_{1}) p_{0} (p_{1}.p_{2}) \\
 &+ p_{1}^{0} (2 (k.p_{2}) M_{W}^2 + 4 (k.p) (p.p_{2}) + (k.p_{1}) (p.p_{2}) + 2 (k.p) (k.p_{1}) s))) \omega + a^4 e^2 \\
& \times(2 (k.p)(k.p_{2}) M_{W}^2 + (k.p_{1})^2 (p_{1}.p_{2}) +  (k.p) (4 (k.p) + (k.p_{1})) (p.p_{2})) \omega^2\\
&  +|\textbf{a}| |\textbf{q}_{1}|(k.p_{1}) (2 |\textbf{a}| (k.p) (2 ((k.p_{2}) M_{W}^2 + 2 (k.p) (p.p_{2})) \omega + (k.p_{1}) (-(k.p_{2}) p_{0}\\
&  + 2 (k.p (p_{1}^{0} + E_{2}) + (p.p_{2}) \omega)) \cos(\theta) + (a_{1}.p_{1}) (k.p_{1})\omega (-2 (k.p) p_{0} + a^2 e^2 \omega)\\
& \times \sin(\theta))\big],
\end{split}
\end{equation}
\begin{equation}
\begin{split}
  C=&\dfrac{e^2}{2 (k.p) (k.p_{1})^3 M_{W}^2}\big[ -4 (k.p) (k.p_{1})^2 (-(a_{1}.p_{1}) (a_{1}.p_{2}) (k.p) +  a^2 ((k.p) (p_{1}.p_{2}) \\
  &+ (k.p_{2}) (M_{W}^2 - (p.p_{1})) + 2 (k.p) (p.p_{2}) + (k.p_{1}) (p.p_{2})))-a^2 (k.p_{1})(a^2 e^2  ((k.p) \\
  &\times (k.p_{2})p_{0} - (k.p_{1}) (k.p_{2}) p_{1}^{0} - 2 (k.p)^2 E_{2}) + 2 (k.p) ((k.p_{1}) p_{0} (p_{1}.p_{2}) +  p_{1}^{0} \\ 
 &\times(2 (k.p_{2}) M_{W}^2 + 4 (k.p) (p.p_{2}) + (k.p_{1})(p.p_{2}) + 2 (k.p) (k.p_{1}) s))) \omega + a^4 e^2 (2 (k.p)\\
 &\times (k.p_{2})   M_{W}^2 + (k.p_{1})^2 (p_{1}.p_{2}) + (k.p) (4 (k.p) + (k.p_{1})) (p.p_{2})) \omega^2 + |\textbf{a}| |\textbf{q}_{1}|(k.p_{1}) \\
 &\times(2 |\textbf{a}| (k.p)(2 ((k.p_{2}) M_{W}^2 + 2 (k.p) (p.p_{2})) \omega + (k.p_{1}) (-(k.p_{2}) p_{0} + 2 (k.p) (p_{1}^{0} \\
 &+ E_{2}) +  (p.p_{2}) \omega)) \cos(\theta) + (a_{1}.p_{1}) (k.p_{1}) \omega (-2 (k.p) p_{0} + a^2 e^2 \omega) \sin(\theta) \big],
 \end{split}
\end{equation}
\begin{equation}
\begin{split}
D=&\dfrac{ e }{(k.p)^2 (k.p_{1})^2 M_{W}^2} \big[  2 (k.p) (k.p_{1}) (-a^2 e^2 (k.p_{1}) (2 (a_{1}.p_{2}) (k.p) + (a_{1}.p_{2}) (k.p_{1}) \\
&+ (a_{1}.p_{1}) (k.p_{2})) + 2 (k.p) (-(a_{1}.p_{2}) (k.p_{1}) M_{W}^2 + (a_{1}.p_{1})(k.p_{2}) M_{W}^2 \\
&+ (a_{1}.p_{2}) (k.p_{1}) (p.p_{1}) + (a_{1}.p_{1}) (2 (k.p) + (k.p_{1})) (p.p_{2}))) + |\textbf{a}| |\textbf{q}_{1}|\omega \\
&\times(a^4 e^4 ((k.p) (2 (k.p) + (k.p_{1})) - 2 (k.p_{1}) (k.p_{2})) \omega -  4 (k.p)^2 (k.p_{1}) (p_{0} (2 (p.p_{2}) \\
&+ (k.p_{1}) s) + M_{W}^2 (p_{1}^{0} + - s \omega)) + 2 a^2 e^2 (k.p) (-(k.p) (k.p_{1}) (p_{0} + 2 (p_{1}^{0} + E_{2})) \\
&+ (k.p) (M_{W}^2 + 2 (k.p_{1}) s) \omega +  (k.p_{1}) (2 (k.p_{2}) p_{0} + 2 (p.p_{2}) \omega + (k.p_{1}) s \omega))) \sin(\theta) \big],
\end{split}
\end{equation}
 \begin{equation}
\begin{split}
E=&\dfrac{e}{(k.p)^2 (k.p_{1})^2 M_{W}^2}\big[ 2 (k.p) (k.p_{1}) (-a^2 e^2 (k.p_{1}) (2 (a_{1}.p_{2}) (k.p) + (a_{1}.p_{2}) (k.p_{1}) \\
&+ (a_{1}.p_{1}) (k.p_{2})) + 2 (k.p) (-(a_{1}.p_{2}) (k.p_{1}) M_{W}^2 + (a_{1}.p_{1}) (k.p_{2}) M_{W}^2 \\
&+ (a_{1}.p_{2}) (k.p_{1}) (p.p_{1}) + (a_{1}.p_{1}) (2 (k.p) + (k.p_{1})) (p.p_{2}))) + |\textbf{a}| |\textbf{q}_{1}|\omega \\
&\times(-a^4 e^4 ((k.p) (2 (k.p) + (k.p_{1})) - 2 (k.p_{1}) (k.p_{2}))\omega + 2 a^2 e^2 (k.p) ((k.p_{1}) \\
&\times(-2 (k.p_{2}) p_{0} + (k.p) (p_{0} + 2 (p_{1}^{0} + E_{2}))) - ((k.p) M_{W}^2 + 2 (k.p_{1}) (p.p_{2}) \\
&+ (k.p_{1}) (2 (k.p) + (k.p_{1})) s)\omega) + 4 (k.p)^2 (k.p_{1}) (p_{0} (2 (p.p_{2}) + (k.p_{1}) s)\\
& + M_{W}^2 (p_{1}^{0} + E_{2} - s \omega))) \sin(\theta)\big],
\end{split}
\end{equation}
\begin{equation}
\begin{split}
F&=\dfrac{4 e^2(a_{1}.p_{1}) (a_{1}.p_{2})  (k.p)}{(k.p_{1}) M_{W}^2},
\end{split}
\end{equation}
where the different scalar products are evaluated in the rest frame of $W^{-}$-boson, and
\begin{equation}
\begin{split}
p_{0}&=M_{W}^{*}+\frac{e^{2}a^{2}\omega}{2(k.p)},\\
p_{1}^{0}&=Q_{1}+\frac{e^{2}a^{2}\omega}{2(k.p_{1})}.
\end{split}
\end{equation}
The expression of $|\textbf{q}_{1}|$ is obtained by solving numerically the equation $g(|\textbf{q}_{1}|)=0$, which is
a required condition to apply the familiar formula given in (\ref{familiarformula}). We mention here that the result obtained in the trace calculation contains the various antisymmetric tensors,  $\epsilon(a,b,c,d)=\epsilon_{\mu\nu\rho\sigma}a^{\mu}b^{\nu}c^{\rho}d^{\sigma}$ for all four-vectors $a, b, c$ and $d$, which arise when four matrices $\gamma$ meet the fifth matrix $\gamma_{5}$ inside  the trace $(\text{Tr}[\gamma^{\mu}\gamma^{\nu}\gamma^{\rho}\gamma^{\sigma}\gamma^{5}]=-4i\epsilon^{\mu\nu\rho\sigma})$. Because it was calculated analytically and replaced in our computational program, it does not appear in the result listed above. We give here an example showing how we calculate these tensors using Einstein's summation and the Grozin convention $\epsilon_{0123}=1$,
\begin{equation}
\begin{split}
\epsilon(a_{1},a_{2},p_{1},p_{2})&=\epsilon_{\mu\nu\rho\sigma}a_{1}^{\mu}a_{2}^{\nu}p_{1}^{\rho}p_{2}^{\sigma},\\
&=|\text{\textbf{a}}|^{2}\big[\epsilon_{1203}p_{1}^{0}p_{2}^{3}+\epsilon_{1230}p_{1}^{3}p_{2}^{0}\big],\\
&=|\text{\textbf{a}}|^{2}\Big[p_{1}^{0}\big(s\omega-|\textbf{q}_{1}|\cos(\theta)\big)-\Big(|\textbf{q}_{1}|\cos(\theta)+\frac{e^{2}a^{2}\omega}{2(k.p_{1})}\Big)p_{2}^{0}\Big].
\end{split}
\end{equation}
So far, we have given a detailed theoretical calculation in order to obtain the expression of the leptonic decay rate of the $W^{-}$-boson in the presence of an EM field. The most known and important quantity that comes directly after calculating the decay rate is the lifetime, and since we only took into account the leptonic decay without the hadronic, we are not allowed to talk about the lifetime of the boson $W^{-}$, because this quantity requires the expression of the total decay rate that combines the leptonic and hadronic $W^{-}$ decays. This is in our perspectives and will be the subject of an upcoming research in which we aim to study the hadronic decay of the boson $W^{-}$ in the presence of an EM field, and then we combine the results obtained with the present leptonic results so that we can explore the effect of the laser on the lifetime as a very important point that must be addressed.
\section{Results and discussion}\label{sec:results}
This section is devoted to the presentation, analysis, and discussion of the numerical results obtained in relation to the leptonic decay rate of the $W^{-}$ boson in the presence of a circularly polarized EM field. Through the following, we will see the effect that the laser may have in terms of its parameters (both its field strength and frequency) on the leptonic decay rate. We chose to have the reference origin on the boson $W$ which is at rest before decay. Let us recall here that the geometry we have worked on inside is a general geometry whose spherical coordinates are $\theta$ and $\varphi$. The angle $\varphi$ was chosen to be zero $(\varphi=0^{\circ})$ in all of the results obtained.
All the results below have been obtained for the following input parameters:
\begin{equation}
\begin{split}
M_{W}&=80.379~\text{GeV},~~ (\text{from \cite{pdg2020}})\\
m_{e}&=0.511\times 10^{-3}~\text{GeV},~~m_{\mu}=0.1057~\text{GeV},~~m_{\tau}=1.777~\text{GeV},\\
e&=-8.5424546\times 10^{-2}~~(\text{charge of electron in natural units}).
\end{split}
\end{equation}
Remember that we have utilized the following unit transformation: in System International (SI) units, the electric field strength $\mathcal{E}_{0}[\text{SI}]=4329.0844~[\text{V/cm}]$ corresponds, in natural units (NU), to $\mathcal{E}_{0}[\text{NU}]=1~[\text{eV}^{2}]$. The first thing we are familiarized to checking in such processes occurring within the EM field is to make sure that the decay rate in the presence of the laser field is exactly equal to its corresponding one in the absence of the laser when we take the zero field strength $(\mathcal{E}_{0}=0~\text{eV})$ and without the exchange of any photons $(s=0)$. In this case, the argument of the Bessel functions is equal to $z=0$. Thus all the terms that contribute to  the "spinorial part" $|\overline{\mathcal{M}^{s}_{fi}}|^{2}$ given in (\ref{result}) will be null except for the term multiplied by the coefficient $A$, since $J_{0}(0)=1$ and $J_{1}(0)=J_{-1}(0)=0$. The expression of $|\overline{\mathcal{M}^{s}_{fi}}|^{2}$ becomes $|\overline{\mathcal{M}^{s}_{fi}}|^{2}=AJ_{s}^{2}(z)/3$, and if we take $a^{2}=0$ we get the trace expression of leptonic $W^{-}$ decay in the absence of the laser field \cite{greiner}; that is
\begin{equation}
|\overline{\mathcal{M}^{laser-free}_{fi}}|^{2}=\frac{8}{3}\bigg((p_{1}.p_{2})+\frac{2(p.p_{1})(p.p_{2})}{M_{W}^{2}}\bigg).
\end{equation}
For the other quantities that compose the decay rate in equation~(\ref{ws}), we finds at $\mathcal{E}_{0}=0$ and $s=0$ their expressions in the absence of the laser field. 
\begin{figure}[h!]
\centering
  \begin{minipage}[t]{0.478\textwidth}
  \centering
    \includegraphics[width=\textwidth]{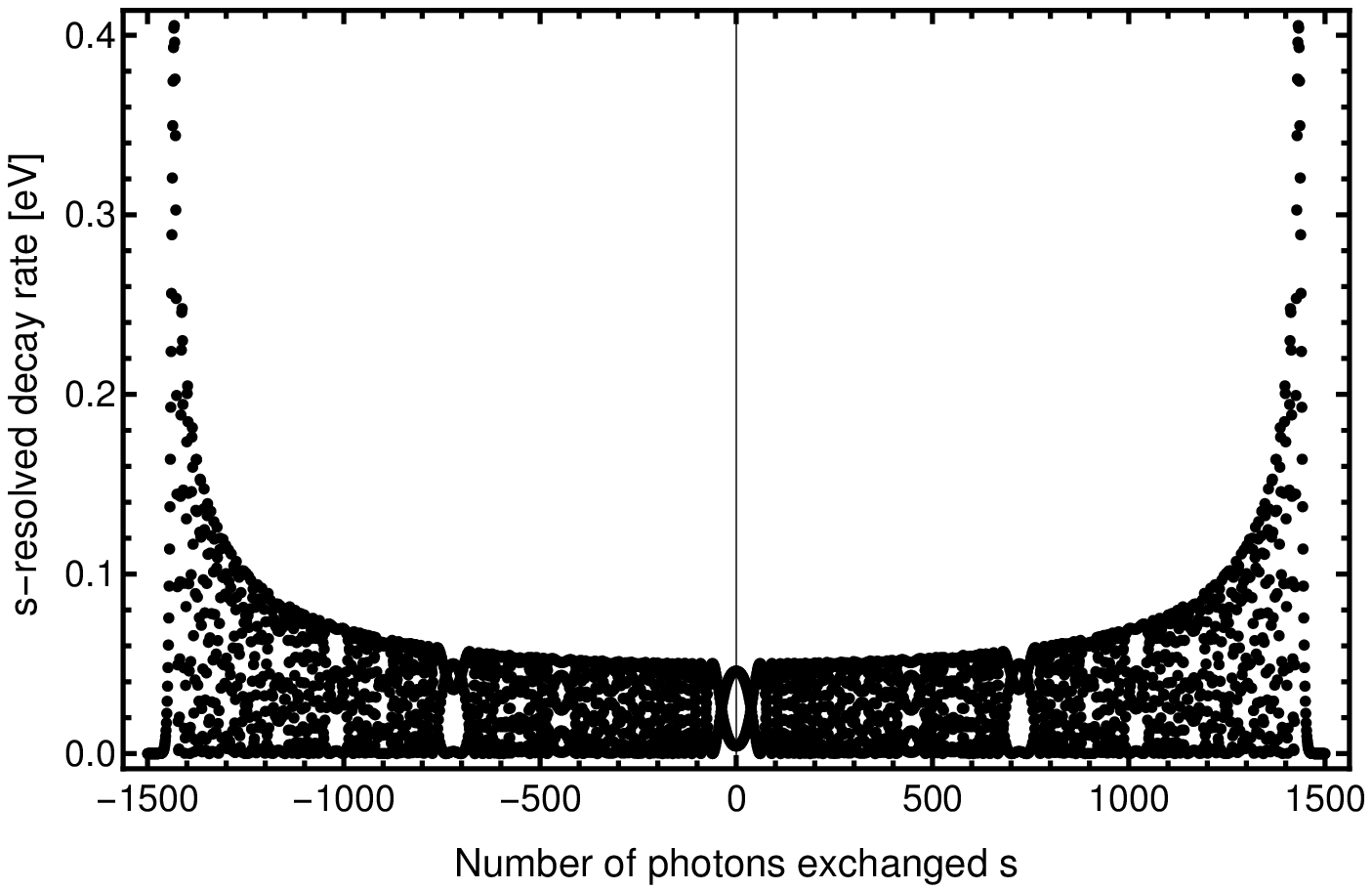}
    \caption{Multiphoton decay rate $d\Gamma^{s}/d\theta~(W^{-}\rightarrow e^{-}\bar{\nu}_{e})$  (\ref{ws}) (in units of $10^{6}$) as a function of the number of photons exchanged $s$ in the rest frame of the $W$-boson, with the spherical coordinates $\theta=90^{\circ}$ and $\varphi=0^{\circ}$. The laser field strength and frequency are $\mathcal{E}_{0}=10^{6}~\text{V/cm}$ and $\hbar\omega=0.117~\text{eV}$ or $\mathcal{E}_{0}=10^{8}~\text{V/cm}$ and $\hbar\omega=1.17~\text{eV}$.}
    \label{fig2}
  \end{minipage}
  \hspace*{0.5cm}
  \begin{minipage}[t]{0.47\textwidth}
  \centering
    \includegraphics[width=\textwidth]{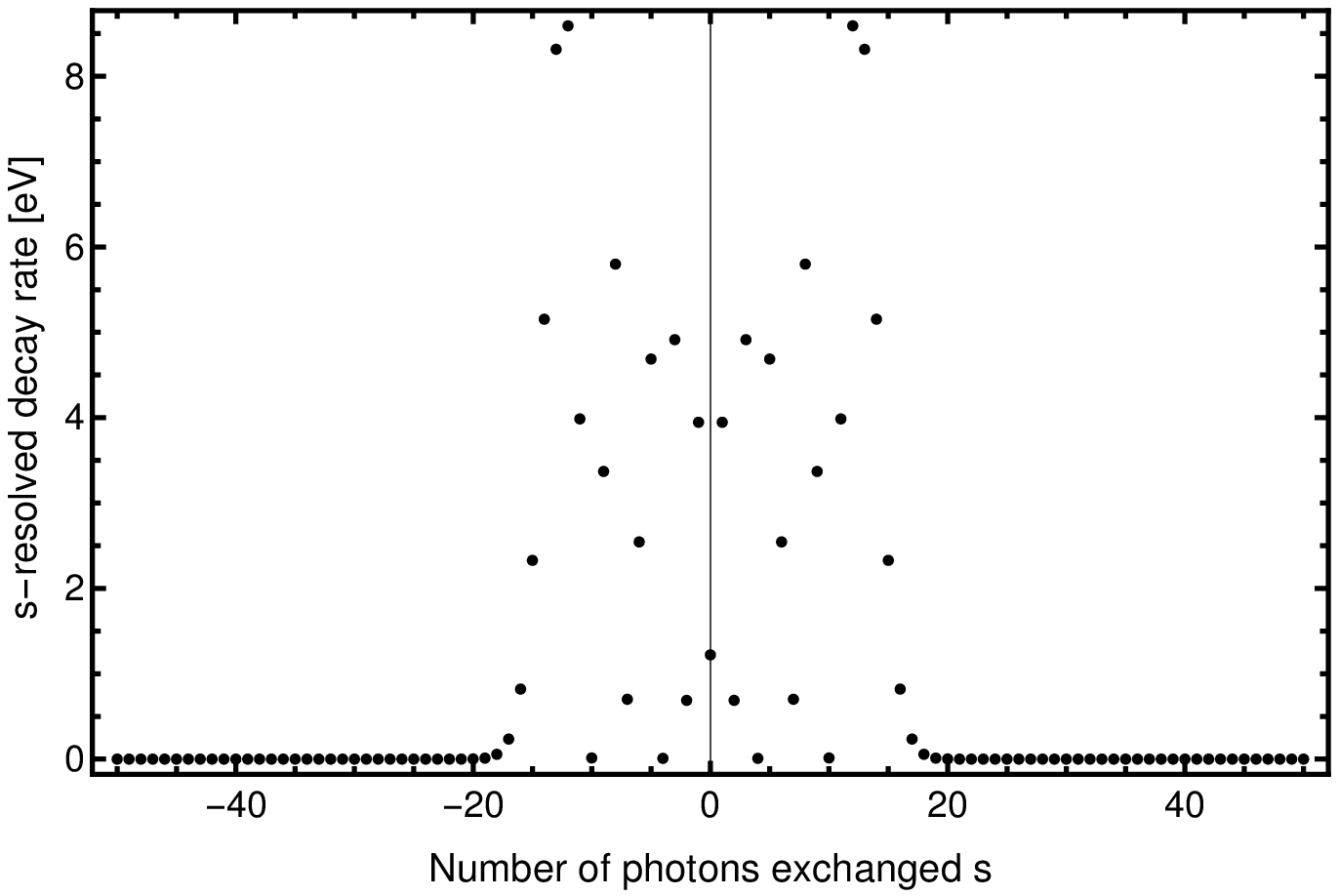}
    \caption{The same as figure \ref{fig2}, but $\mathcal{E}_{0}=10^{6}~\text{V/cm}$ and $\hbar\omega=1.17~\text{eV}$, }
    \label{fig3}
  \end{minipage}
   \begin{minipage}[t]{0.478\textwidth}
  \centering
    \includegraphics[width=\textwidth]{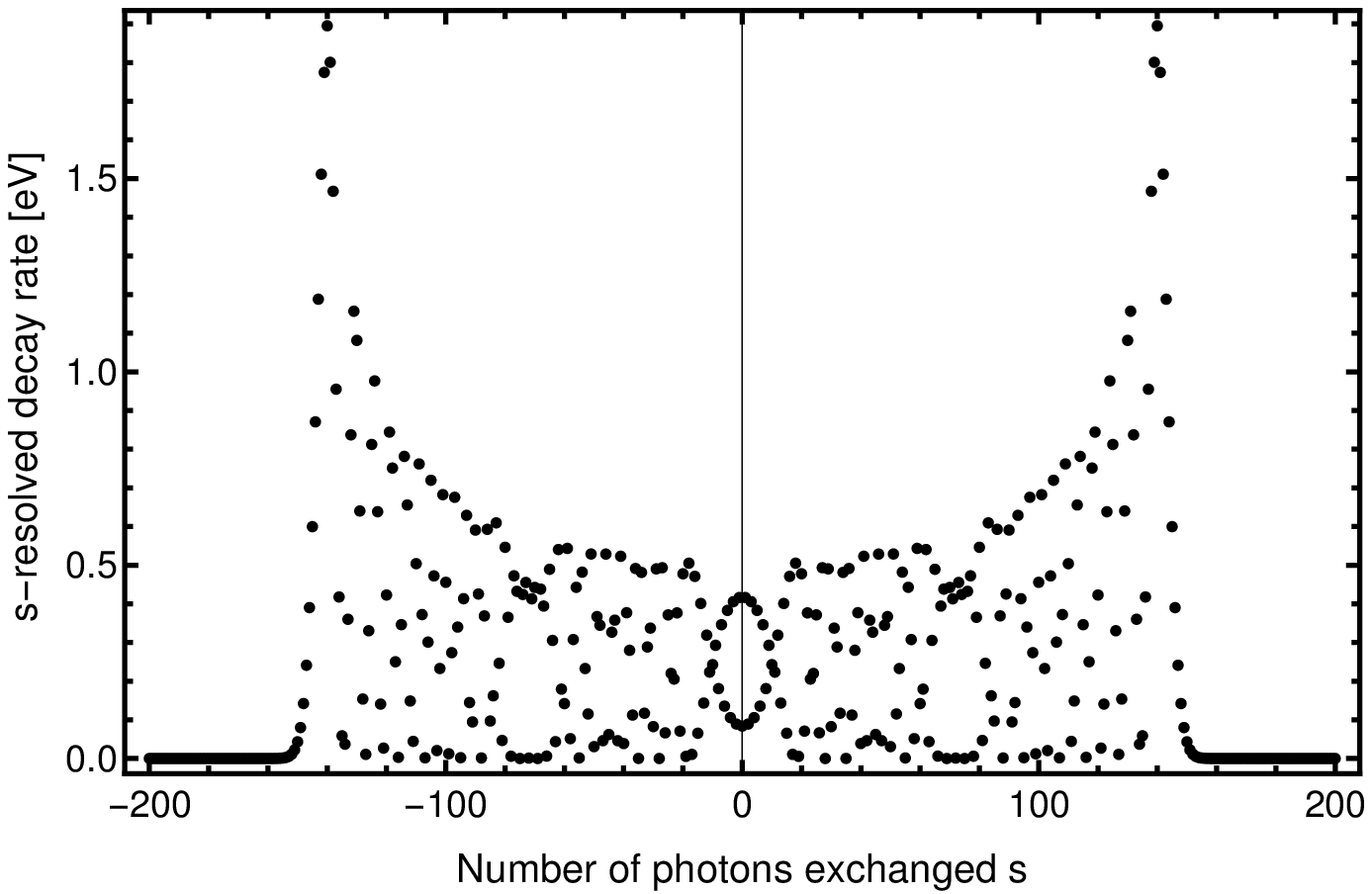}
    \caption{The same as figure \ref{fig2}, but $\mathcal{E}_{0}=10^{7}~\text{V/cm}$ and $\hbar\omega=1.17~\text{eV}$,}
    \label{fig4}
  \end{minipage}
  \hspace*{0.5cm}
  \begin{minipage}[t]{0.47\textwidth}
  \centering
    \includegraphics[width=\textwidth]{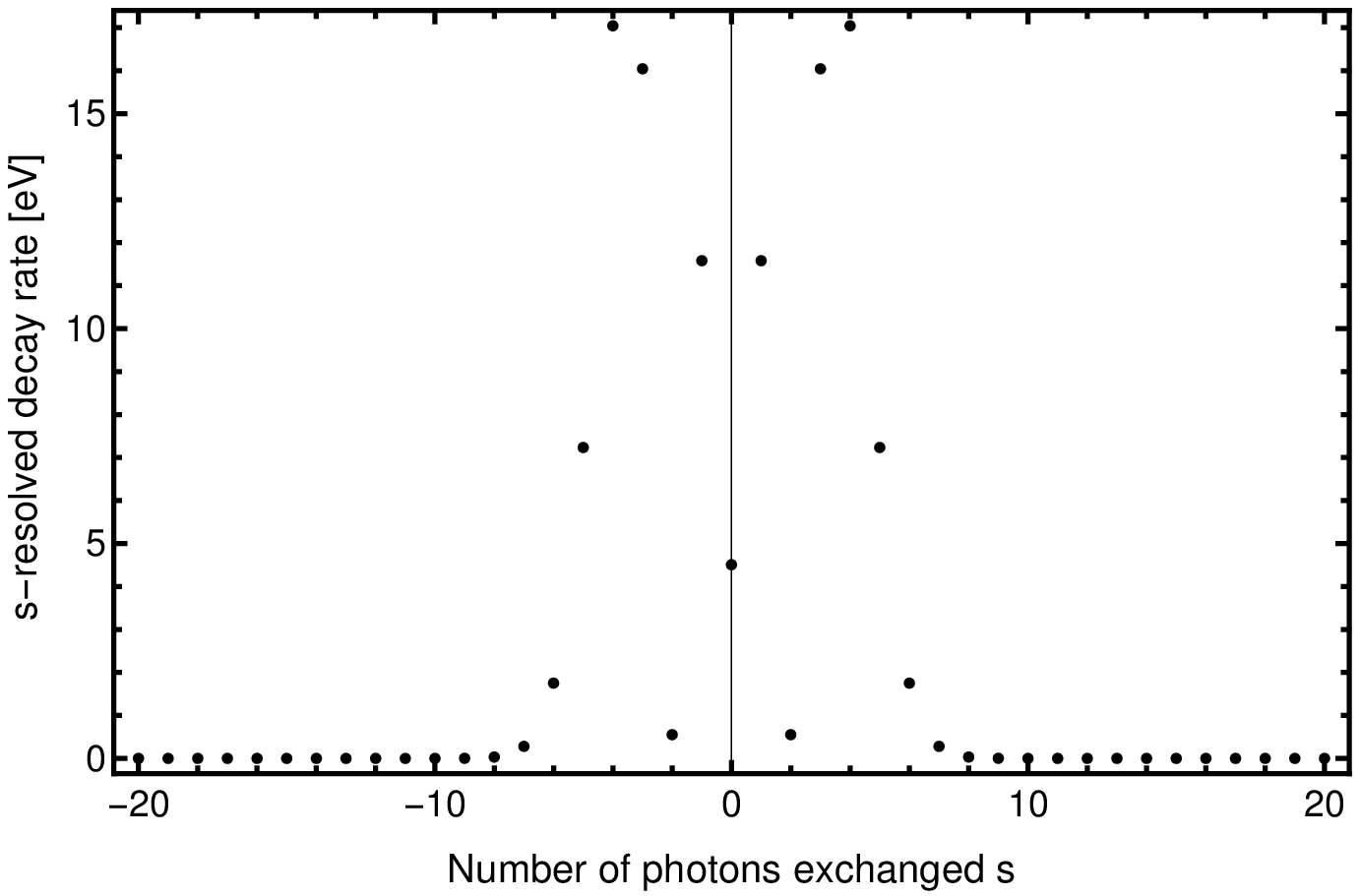}
    \caption{The same as figure \ref{fig2}, but $\mathcal{E}_{0}=10^{6}~\text{V/cm}$ and $\hbar\omega=2~\text{eV}$.}
    \label{fig5}
  \end{minipage}
\end{figure}
%\begin{figure}[hbtp]
%\centering
%  \subcaptionbox{\label{a}}{\includegraphics[height=5cm,width=.47\linewidth]{fig2}}\hspace*{0.5cm}
%  \subcaptionbox{\label{b}}{\includegraphics[height=5cm,width=.47\linewidth]{fig3}}\par
%  \subcaptionbox{\label{c}}{\includegraphics[height=5cm,width=.47\linewidth]{fig4}}\hspace*{0.5cm}
%  \subcaptionbox{\label{d}}{\includegraphics[height=5cm,width=.47\linewidth]{fig5}}
%  \caption{Multiphoton decay rate $d\Gamma^{s}/d\theta~(W^{-}\rightarrow e^{-}\bar{\nu}_{e})$  (\ref{ws}) (in units of $10^{6}$) as a function of the number of photons exchanged $s$ in the rest frame of the $W$-boson, with the spherical coordinates $\theta=90^{\circ}$ and $\varphi=0^{\circ}$. The laser field strength and frequency are (a) $\mathcal{E}_{0}=10^{6}~\text{V/cm}$ and $\hbar\omega=0.117~\text{eV}$ or $\mathcal{E}_{0}=10^{8}~\text{V/cm}$ and $\hbar\omega=1.17~\text{eV}$, (b) $\mathcal{E}_{0}=10^{6}~\text{V/cm}$ and $\hbar\omega=1.17~\text{eV}$, (c) $\mathcal{E}_{0}=10^{7}~\text{V/cm}$ and $\hbar\omega=1.17~\text{eV}$, and (d) $\mathcal{E}_{0}=10^{6}~\text{V/cm}$ and $\hbar\omega=2~\text{eV}$.}\label{fig2}
%\end{figure}
\\
Figures \ref{fig2}-\ref{fig5} show the multiphoton decay rate $d\Gamma^{s}/d\theta$ (\ref{ws}) for the process $(W^{-}\rightarrow e^{-}\bar{\nu}_{e})$ versus the net photon number $s$ transferred between the decaying system and the laser field. In order to show numerically the effect of field strength and frequency on the phenomenon of multiphoton energy transfer, we display the s-resolved decay rate at different field strengths and frequencies for the angle $\theta=90^{\circ}$. To examine the effect of the laser field strength in the first step, we set, in figures \ref{fig2}, \ref{fig3} and \ref{fig4} the frequency to the value $\hbar\omega=1.17~\text{eV}$ and changed the field strength $\mathcal{E}_{0}$ so that it is equal to $\mathcal{E}_{0}=10^{6}~\text{V/cm}$ in figure \ref{fig3}, $\mathcal{E}_{0}=10^{7}~\text{V/cm}$ in figure \ref{fig4} and $\mathcal{E}_{0}=10^{8}~\text{V/cm}$ in figure \ref{fig2}. Comparing these three figures with each other, we can see that as the laser field strength increases at a fixed frequency, the number of photons exchanged enhances and increases, and this is evident through the value of the cutoff number in each case. The cutoff numbers in figures \ref{fig3}, \ref{fig4} and \ref{fig2} are, respectively, $20$, $160$ and $1500$ (for the absorption side) and $-20$, $-160$ and $-1500$ (for the emission side). It can also be observed from these figures that the magnitude of $d\Gamma^{s}/d\theta$ decreases as the field strength increases. This is the effect of laser field strength. For the effect of laser frequency, we have exactly reversed the previous situation. In figures \ref{fig2}, \ref{fig3} and \ref{fig5}, we fixed the field strength to the value $\mathcal{E}_{0}=10^{6}~\text{V/cm}$ and changed the laser frequency from $\hbar\omega=0.117~\text{eV}$ in figure \ref{fig2} and $\hbar\omega=1.17~\text{eV}$ in figure \ref{fig3} to the value $\hbar\omega=2~\text{eV}$ in figure \ref{fig5}. Through these figures, we can see that, unlike the field strength effect, when the laser frequency increases, the number of exchanged photons decreases, and this is evident by the decrease of the cutoff number from the value $1500$ at frequency $\hbar\omega=0.117~\text{eV}$ [figure \ref{fig2}] to $10$ at frequency $\hbar\omega=2~\text{eV}$ [figure \ref{fig5}] with a constant field strength equal to $\mathcal{E}_{0}=10^{6}~\text{V/cm}$. In relation to the magnitude of  $d\Gamma^{s}/d\theta$, it is also noted that it increases with the increase of the laser frequency until reaching, for example, $d\Gamma^{s}/d\theta = 17\times 10^{6}~\text{eV}$ at frequency $\hbar\omega=2~\text{eV}$ and $s=4$ [see figure \ref{fig5}]. Based on the foregoing, it can be said that for increasing field strength and fixed frequency, processes involving large numbers of photons make significant contributions, in contrast to the case of high frequencies and fixed field strength where only low-photon exchanging processes are important. This is demonstrated by the formula of the argument $z=-e(a_{1}.p_{1})/(k.p_{1})\propto \mathcal{E}_{0}/\omega^{2}$ given in (\ref{argument}), which is responsible for the effects of the laser, as it decreases with higher frequencies and vice versa, meaning that at lower frequencies a greater number of photons can be exchanged; as can be deduced from the behavior of the Bessel function $J_{s}(z)$ and clearly seen in figures \ref{fig2}-\ref{fig5}. Thus, for stronger field strengths or lower frequencies, an increased effect on the leptonic decay rate can be expected. The heights of the different photon-number peaks are critically dependent on the values of the ordinary Bessel functions. In all the previous results presented in figures \ref{fig2}-\ref{fig5}, the contributions of the different $s$-photon processes are cut off at two edges that are symmetric with respect to $s = 0$. It also appears that all the spectra are symmetric envelopes with respect to both the positive and negative edges, and this indicates that the photon absorption  processes $(s>0)$ are exactly equal to the photon emission processes $(s<0)$. The cutoff number can be explained by the well-known properties of the Bessel function, which decreases considerably when its argument $z$ (equation \ref{argument}) is approximately equal to its order $s$. This has already been referred to in refs. \cite{szymanowski,liberakdar}. In fact, our goal in drawing these envelopes in terms of the number of exchanged photons is to obtain the cutoff numbers at each specified field strength and frequency. The cutoff number is defined as a fixed number of photons from which the s-resolved decay rate $d\Gamma^{s}/d\theta$ suddenly falls off $(d\Gamma^{s}/d\theta=0)$. This number is very important to us because with it we make sure that the famous Kroll–Watson sum rule \cite{sumrule} is fulfilled. By the way, and in this context, we mention here that this sum rule in our case can be formulated mathematically as follows:
\begin{equation}
\sum_{s=-\text{cutoff}}^{+\text{cutoff}} \Gamma^{s}(W^{-}\rightarrow \ell^{-}\bar{\nu}_{\ell})=\Gamma^{\text{laser-free}}(W^{-}\rightarrow \ell^{-}\bar{\nu}_{\ell}),
\end{equation}
and is achieved when the decay rate in the presence of the laser field tends to approach the laser-free results by increasing the number of exchanged photons until they are exactly equal when we reach the cutoff number. In the next stage, we will try to make sure that this sum rule is respected. Let us take for example the case presented in figure \ref{fig4} where the field strength $\mathcal{E}_{0}=10^{7}~\text{V/cm}$ and frequency $\hbar\omega=1.17~\text{eV}$, the cutoff number in this case is equal, as shown in figure \ref{fig4}, to about $+160$ (for absorption) and $-160$ (for emission). In reality, this number of exchanged photons cannot be considered a cutoff number according to the aforementioned definition, because the decay rate $d\Gamma^{s}/d\theta$ is not exactly equal to $0$; but it is so small that our numerical program gives it a zero value when drawing the curves. By checking the values numerically, we found that the decay rate $d\Gamma^{s}/d\theta$ is zero exactly at $s\geq450$ ($s\leq -450$). Therefore, the sum rule must be satisfied essentially at this number of exchanged photons. To prove this, we plotted the decay rate $\Gamma(W^{-}\rightarrow e^{-}\bar{\nu}_{e})$ (\ref{summed}), at $\mathcal{E}_{0}=10^{7}~\text{V/cm}$ and $\hbar\omega=1.17~\text{eV}$, summed over an increasingly different numbers of exchanged photons and compared it with the decay rate in the absence of the laser field. The curves obtained in connection with this are shown in figure \ref{fig6}, where the numbers of exchanged photons over which we have summed the leptonic decay rate are as follows: $-N\leq s \leq +N$, where $N=30,~60,~120,~180,~200,~450$. Through this figure, it is evident to us that with the increase in the number of photons exchanged between the decaying system and the laser field, the effect of the latter on the leptonic decay rate gradually decreases until it becomes completely absent when the number of exchanged photons is equal to the cutoff number $(-450\leq s \leq +450$ in this case$)$, and thus we obtain two identical curves. In other words, we say that the leptonic decay rate summed over a number of exchanged photons approaches slowly and tends, as the number of exchanged photons increases, towards its corresponding one in the absence of the laser until it is exactly equal to it at a specified number of exchanged photons, which is the cutoff number. This is exactly what we mean by the sum rule which is, in our case, respected and successfully achieved.
\begin{figure}
\centering
  \begin{minipage}[t]{0.47\textwidth}
  \centering
    \includegraphics[width=\textwidth]{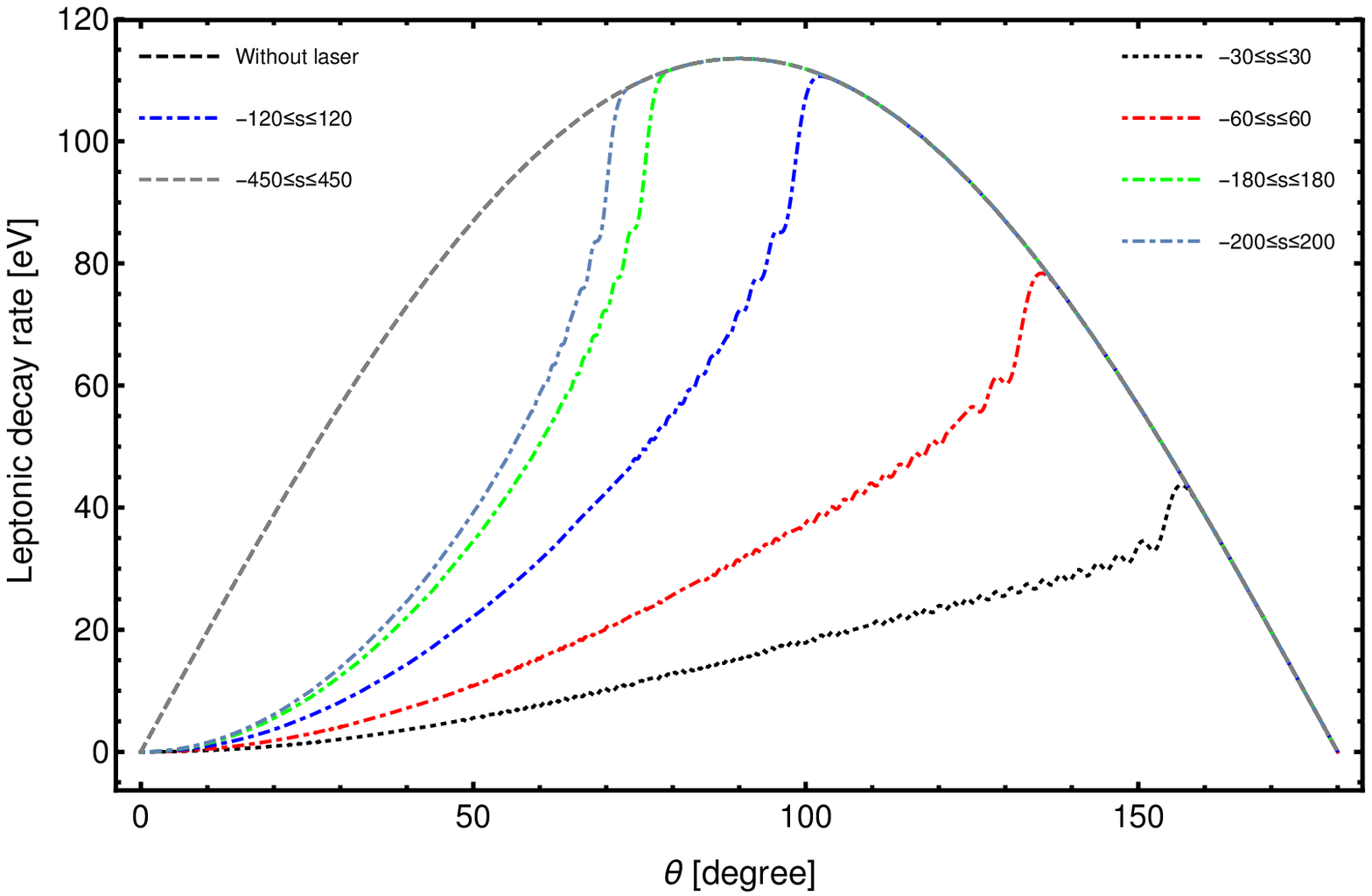}
    \caption{The variation of the summed leptonic decay rate $\Gamma(W^{-}\rightarrow e^{-}\bar{\nu}_{e})$ (\ref{summed}) (in units of $10^{6}$) with and without a laser as a function of the angle $\theta$ for various numbers of photons exchanged. The spherical coordinate $\varphi=0^{\circ}$. The laser field strength and frequency are $\mathcal{E}_{0}=10^{7}~\text{V/cm}$ and $\hbar\omega=1.17~\text{eV}$.}
    \label{fig6}
  \end{minipage} %
  \hspace*{0.5cm}
  \begin{minipage}[t]{0.47\textwidth}
  \centering
    \includegraphics[width=\textwidth]{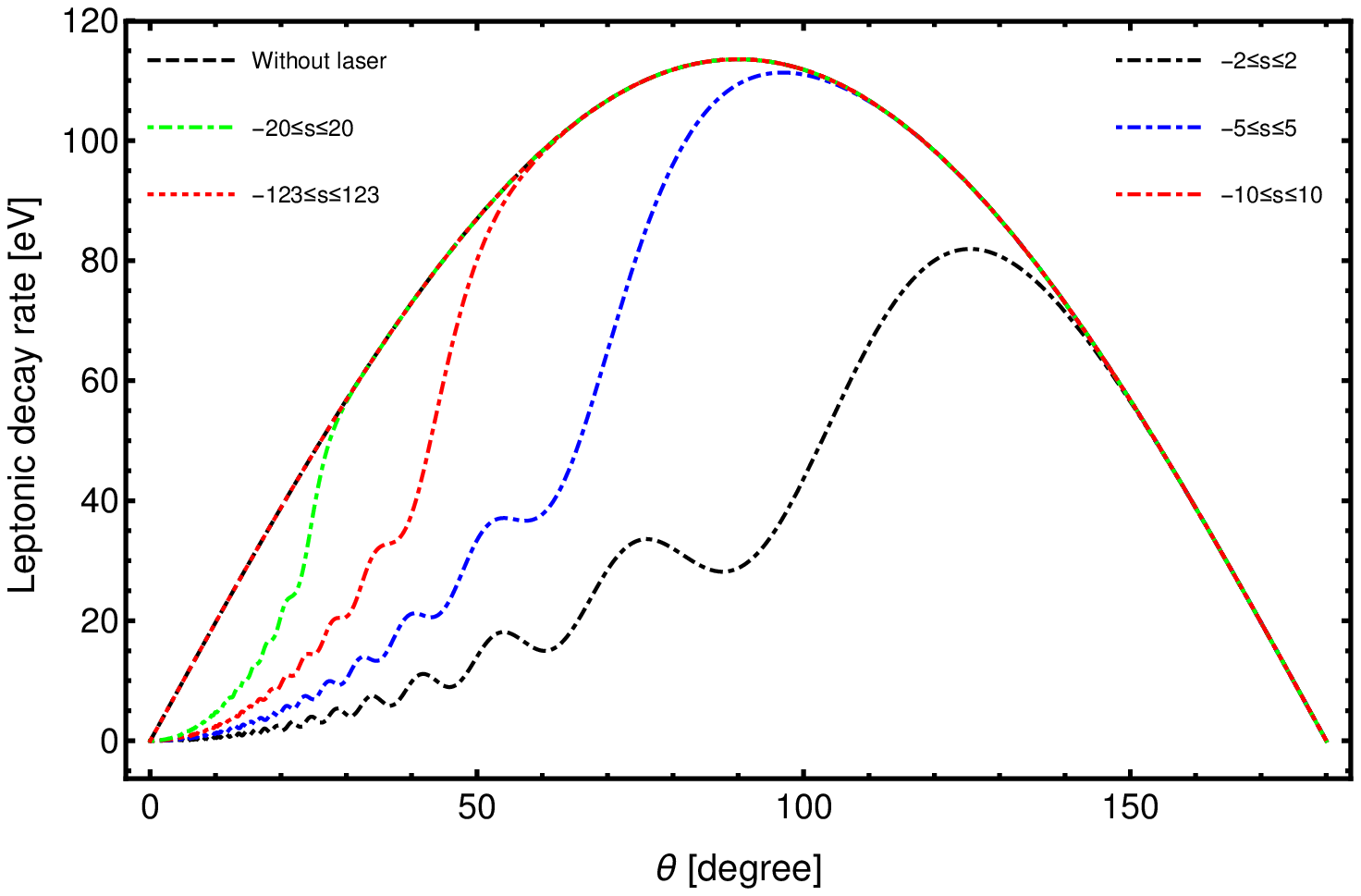}
    \caption{The same as figure \ref{fig6} but for $\mathcal{E}_{0}=10^{6}~\text{V/cm}$ and $\hbar\omega=2~\text{eV}$.}
    \label{fig7}
  \end{minipage}
\end{figure}
%\begin{figure}[hbtp]
%\centering
%  \subcaptionbox{\label{a3}}{\includegraphics[height=5cm,width=.47\linewidth]{fig6}}\hspace*{0.5cm}
%  \subcaptionbox{\label{b3}}{\includegraphics[height=5cm,width=.47\linewidth]{fig7}}
%  \caption{The variation of the summed leptonic decay rate $\Gamma(W^{-}\rightarrow e^{-}\bar{\nu}_{e})$ (\ref{summed}) (in units of $10^{6}$) with and without a laser as a function of the angle $\theta$ for various numbers of photons exchanged. The spherical coordinate $\varphi=0^{\circ}$. The laser field strength and frequency are (a) $\mathcal{E}_{0}=10^{7}~\text{V/cm}$ and $\hbar\omega=1.17~\text{eV}$, (b) $\mathcal{E}_{0}=10^{6}~\text{V/cm}$ and $\hbar\omega=2~\text{eV}$.}\label{fig3}
%\end{figure}
\\
The idea of the sum rule was originally conceived when Kroll and Watson derived, in $1973$, the differential cross-section formula for an elastic electron scattering process in the presence of a laser when neglecting the interaction of the atom (scattering potential) with the laser. Within the framework of the Kroll-Watson approximation, the information obtained from the laser-assisted leptonic decay rate regarding the decay system is the same as the information obtained from the decay rate without the laser field. Figure \ref{fig7} shows the same thing as figure \ref{fig6}, except that in this case the laser field strength and frequency are, respectively, $\mathcal{E}_{0}=10^{6}~\text{V/cm}$ and $\hbar\omega=2~\text{eV}$, which are the same values used for each of them to draw the envelope in figure \ref{fig5}. As shown in figure \ref{fig5}, the cutoff number equals $10$ for absorption ($-10$ for emission). But its real value, from which the s-resolved decay rate is zero, is $123$. Therefore, we see that the sum rule in figure \ref{fig7} is fulfilled exactly at this number of exchanged photons. These are two clear examples that are sufficient to make sure that the well-known sum rule is respected and therefore can be served and considered as a proof of the correctness of our theoretical calculations.
In order to realize the size of the laser effect on the leptonic decay rate and make it more clear, it is convenient to introduce here a quantity denoted by $R_{with/without}$ and defined as the ratio between the summed leptonic decay rate in the EM field given in equation~(\ref{summed}) and its corresponding one in the absence of the laser field; that is,
\begin{equation}\label{ratio}
R_{with/without}=\frac{\Gamma^{with~laser}(W^{-}\rightarrow e^{-}\bar{\nu}_{e})}{\Gamma^{without~laser}(W^{-}\rightarrow e^{-}\bar{\nu}_{e})}.
\end{equation}
\begin{figure}[h!]
\centering
  \begin{minipage}[t]{0.46\textwidth}
  \centering
    \includegraphics[width=\textwidth]{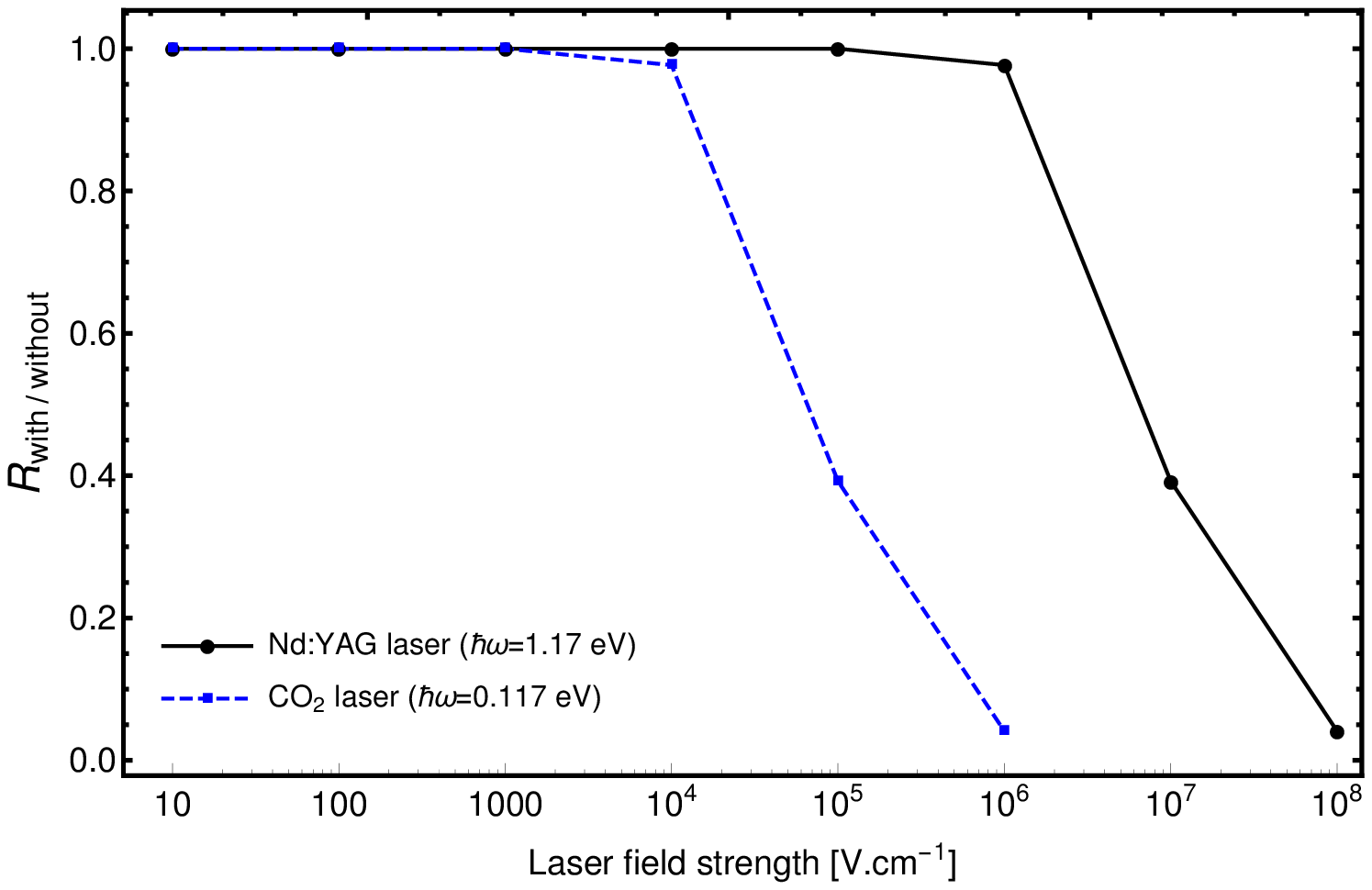}
    \caption{The variation of the ratio $R_{with/without}$ (\ref{ratio}) as a function of the laser field strength for a Nd:YAG laser ($\hbar\omega=1.17~\text{eV}$) and a $\text{CO}_{2}$ laser ($\hbar\omega=0.117~\text{eV}$). The number of exchanged photons summed over it is $-60\leq s\leq+60$.}
    \label{fig8}
  \end{minipage} %
  \hspace*{0.5cm}
  \begin{minipage}[t]{0.48\textwidth}
  \centering
    \includegraphics[width=\textwidth]{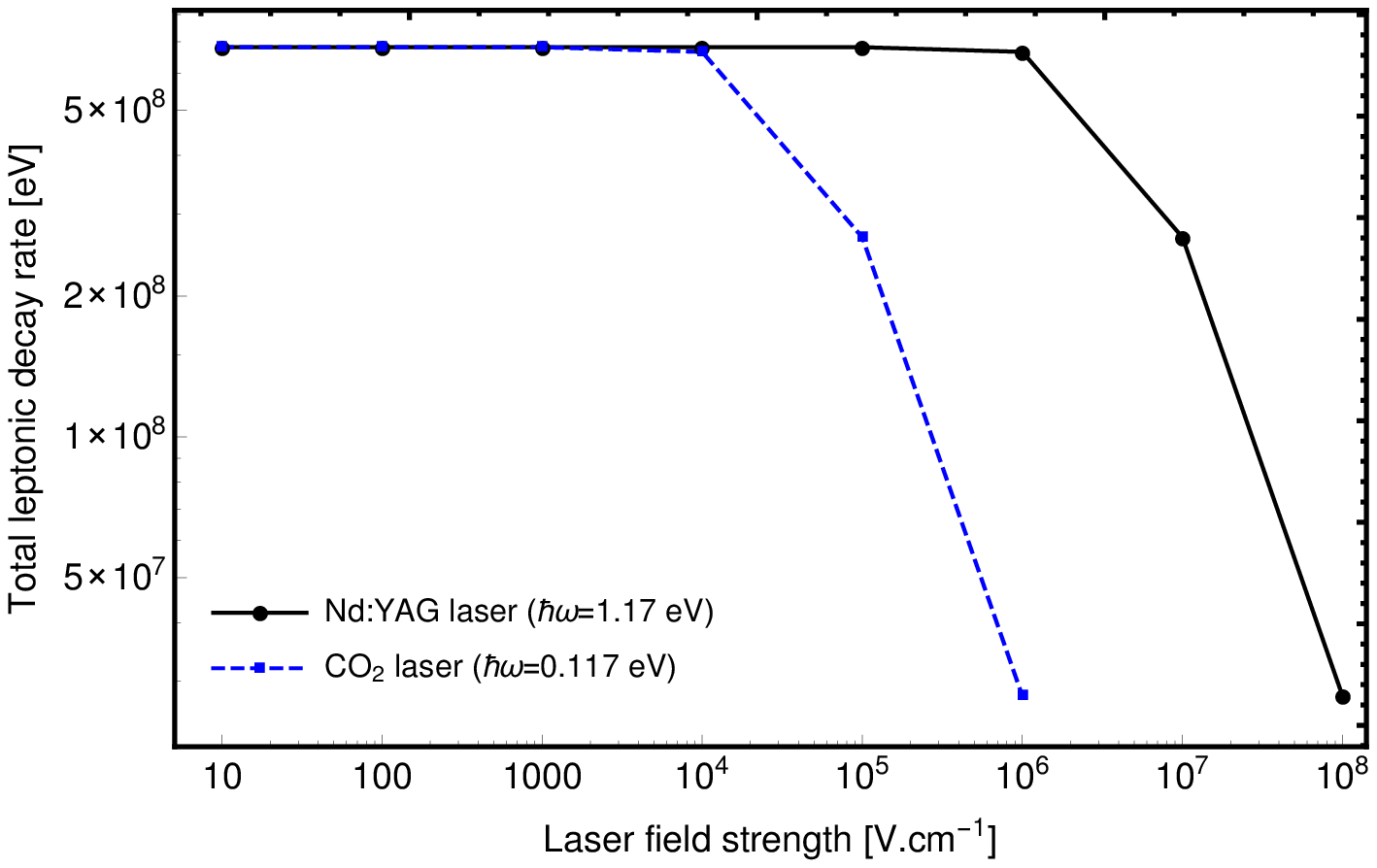}
    \caption{The variation of the total leptonic decay rate (sum over leptons) as a function of the laser field strength for a Nd:YAG laser ($\hbar\omega=1.17~\text{eV}$) and a $\text{CO}_{2}$ laser ($\hbar\omega=0.117~\text{eV}$). The number of exchanged photons summed over it is $-60\leq s\leq+60$.}
    \label{fig9}
  \end{minipage}
\end{figure}
%\begin{figure}[hbtp]
%\centering
%  \subcaptionbox{\label{a4}}{\includegraphics[height=5cm,width=.47\linewidth]{fig8}}\hspace*{0.5cm}
%  \subcaptionbox{\label{b4}}{\includegraphics[height=5cm,width=.47\linewidth]{fig9}}
%  \caption{The variation of (a) the ratio $R_{with/without}$ (\ref{ratio}) and (b) the total leptonic decay rate (sum over leptons) as a function of the laser field strength for a Nd:YAG laser ($\hbar\omega=1.17~\text{eV}$) and a $\text{CO}_{2}$ laser ($\hbar\omega=0.117~\text{eV}$). The number of exchanged photons summed over it is $-60\leq s\leq+60$.}\label{fig4}
%\end{figure}
In figures \ref{fig8} and \ref{fig9}, we aim to illustrate the effect of laser frequency on both the ratio $R_{with/without}$ (\ref{ratio}) and the total leptonic decay rate, which is equal to the sum of the electron, muon and tau decay rates. To do this, we plotted their changes in terms of laser field strength for two different frequencies, $\text{CO}_{2}$ laser frequency ($\hbar\omega=0.117~\text{eV}$) and Nd:YAG laser frequency ($\hbar\omega=1.17~\text{eV}$). Based on figure \ref{fig8}, it can be seen that the $\text{CO}_{2}$ laser affects the $R_{with/without}$ ratio only when the field strength is above the threshold of $10^{3}~\text{V/cm}$, while the effect of the Nd:YAG laser on the ratio starts only when the field strength is above $10^{5}~\text{V/cm}$; so the $\text{CO}_{2}$ laser, whose frequency is lower than the Nd:YAG one, can affect the $R_{with/without}$ ratio more than the Nd:YAG laser even at low field strengths. For example, the $R_{with/without}$ value at $\mathcal{E}_{0}=10^{5}~\text{V/cm}$ in the presence of the $\text{CO}_{2}$ laser $(R_{with/without}=0.4)$ does not attained in the presence of the Nd:YAG laser except when $\mathcal{E}_{0}=10^{7}~\text{V/cm}$. Thus, having a high-frequency laser effect requires the availability of very high field strengths. The same can be observed completely with regard to the effect of the laser frequency on the total leptonic decay rate shown in figure \ref{fig9}. It can be said that the effect of the laser decreases at high frequency or we say that the high-frequency laser has little effect compared to the low-frequency laser.
\begin{figure}
\centering
  \begin{minipage}[t]{0.46\textwidth}
  \centering
    \includegraphics[width=\textwidth]{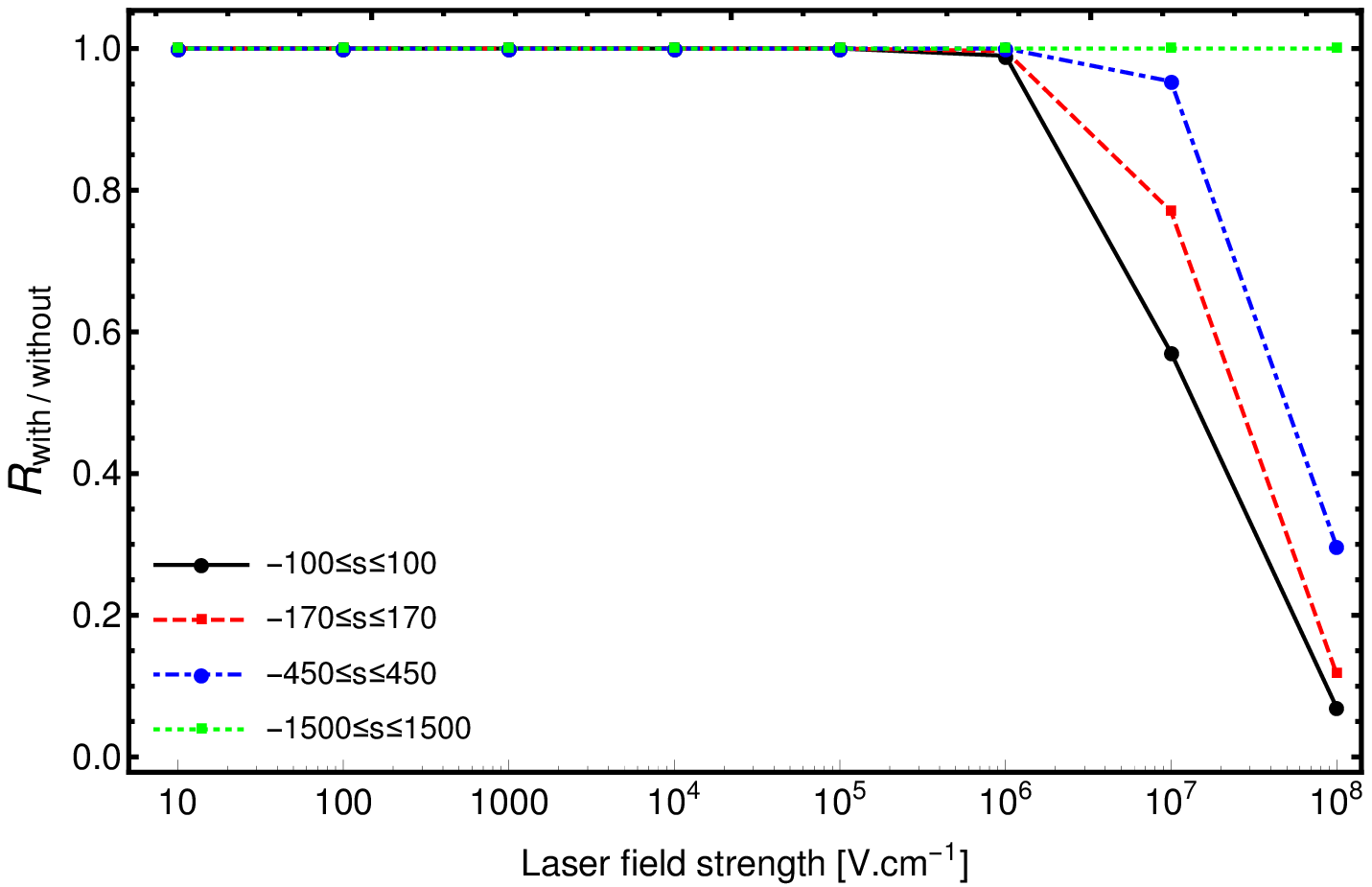}
    \caption{The same as in figure \ref{fig8}, but for different numbers of exchanged photons and fixed laser frequency $\hbar\omega=1.17~\text{eV}$.}
    \label{fig10}
  \end{minipage} %
  \hspace*{0.5cm}
  \begin{minipage}[t]{0.48\textwidth}
  \centering
    \includegraphics[width=\textwidth]{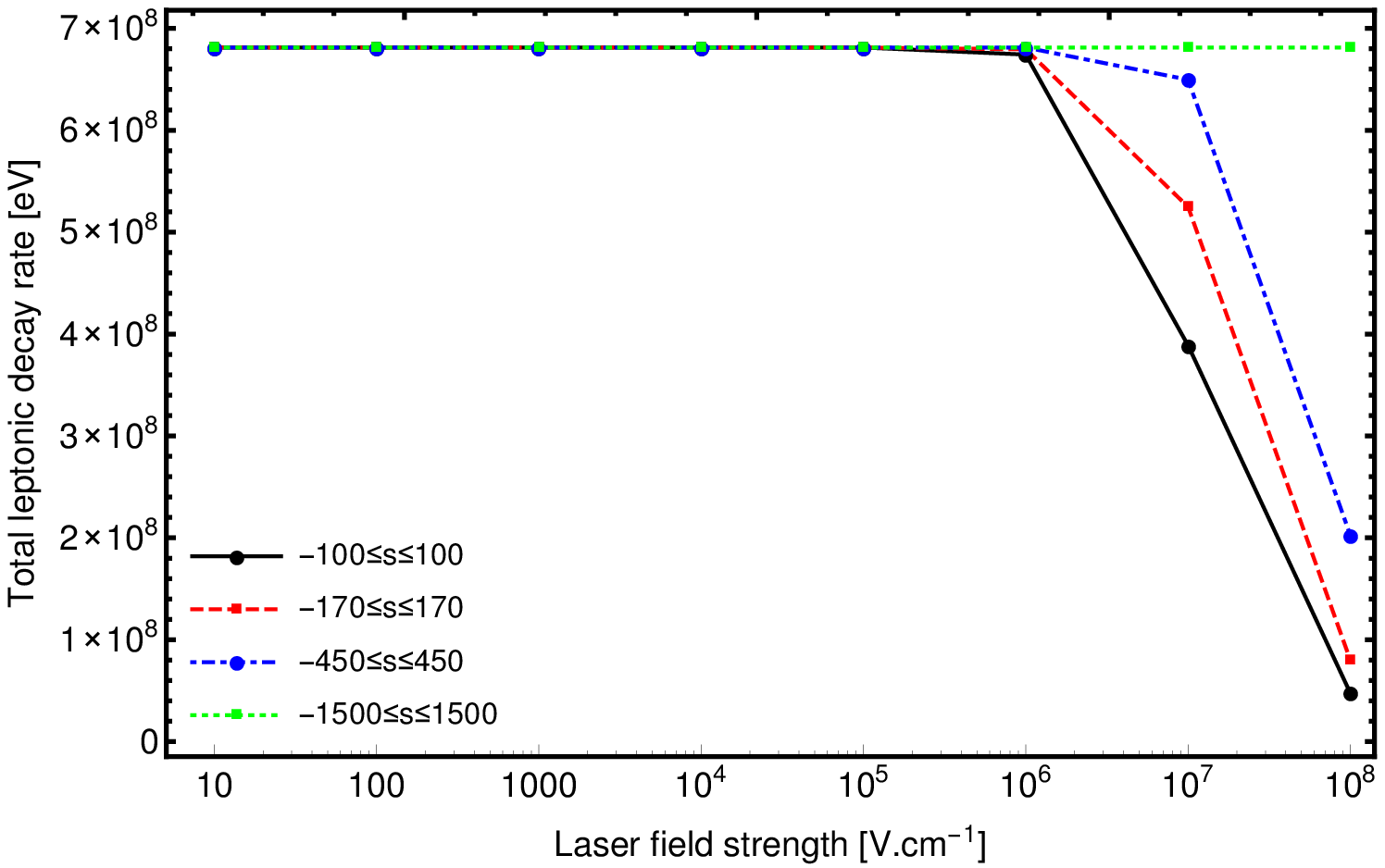}
    \caption{The same as in figure \ref{fig9}, but for different numbers of exchanged photons and fixed laser frequency $\hbar\omega=1.17~\text{eV}$.}
    \label{fig11}
  \end{minipage}
\end{figure}
%\begin{figure}[hbtp]
%\centering
%  \subcaptionbox{\label{a5}}{\includegraphics[height=5cm,width=.47\linewidth]{fig10}}\hspace*{0.5cm}
%  \subcaptionbox{\label{b5}}{\includegraphics[height=5cm,width=.47\linewidth]{fig11}}
%  \caption{The same as in figure \ref{fig4}, but for different numbers of exchanged photons and fixed laser frequency $\hbar\omega=1.17~\text{eV}$.}\label{fig5}
%\end{figure}
In figures \ref{fig10} and \ref{fig11}, we have drawn the same two quantities as in figures \ref{fig8} and \ref{fig9} at different numbers of photons exchanged, but now we have set the laser frequency to $\hbar\omega=1.17~\text{eV}$. This is exactly the same as we did when we wanted to check the sum rule. We will show here via figures \ref{fig10} and \ref{fig11} how the sum rule is also respected in both the $R_{with/without}$ ratio and the total leptonic decay rate. We have summed over $100$ and three cutoff numbers $170$, $450$ and $1500$. We notice that at small field strengths all the curves, regardless of the number of exchanged photons, are coincident and have a constant value of $1$ with respect to $R_{with/without}$ in figure \ref{fig10} and a value $6.81192\times10^{8}~\text{eV}$ for the total leptonic decay rate in the absence of the laser in figure \ref{fig11}. Above the strength $10^{5}~\text{V/cm}$, we notice that both quantities, for each specified number of exchanged photons, decrease nonlinearly with increasing field strength. It appears clearly to us that with the increase in the number of exchanged photons, the laser's influence whether on the ratio or the total decay rate decreases until it becomes completely absent when we reach the cutoff $-1500\leq s\leq +1500$ at which the sum rule is fulfilled, and thus we obtain a constant horizontal curve at the field strengths between $10$ and $10^{8}~\text{V/cm}$. The same thing we have recently examined and verified concerning the lifetime in the case of the pion and $Z$-boson decays \cite{mouslih,jakha}, and we will confirm it in a future study regarding the lifetime of the $W$-boson. The final result that we can deduce from the above is that the effect of the laser on the decay rate remains present as long as the number of exchanged photons that we sum over it does not reach the cutoff number at which the sum rule is fulfilled.
\begin{figure}
\centering
  \begin{minipage}[t]{0.46\textwidth}
  \centering
    \includegraphics[width=\textwidth]{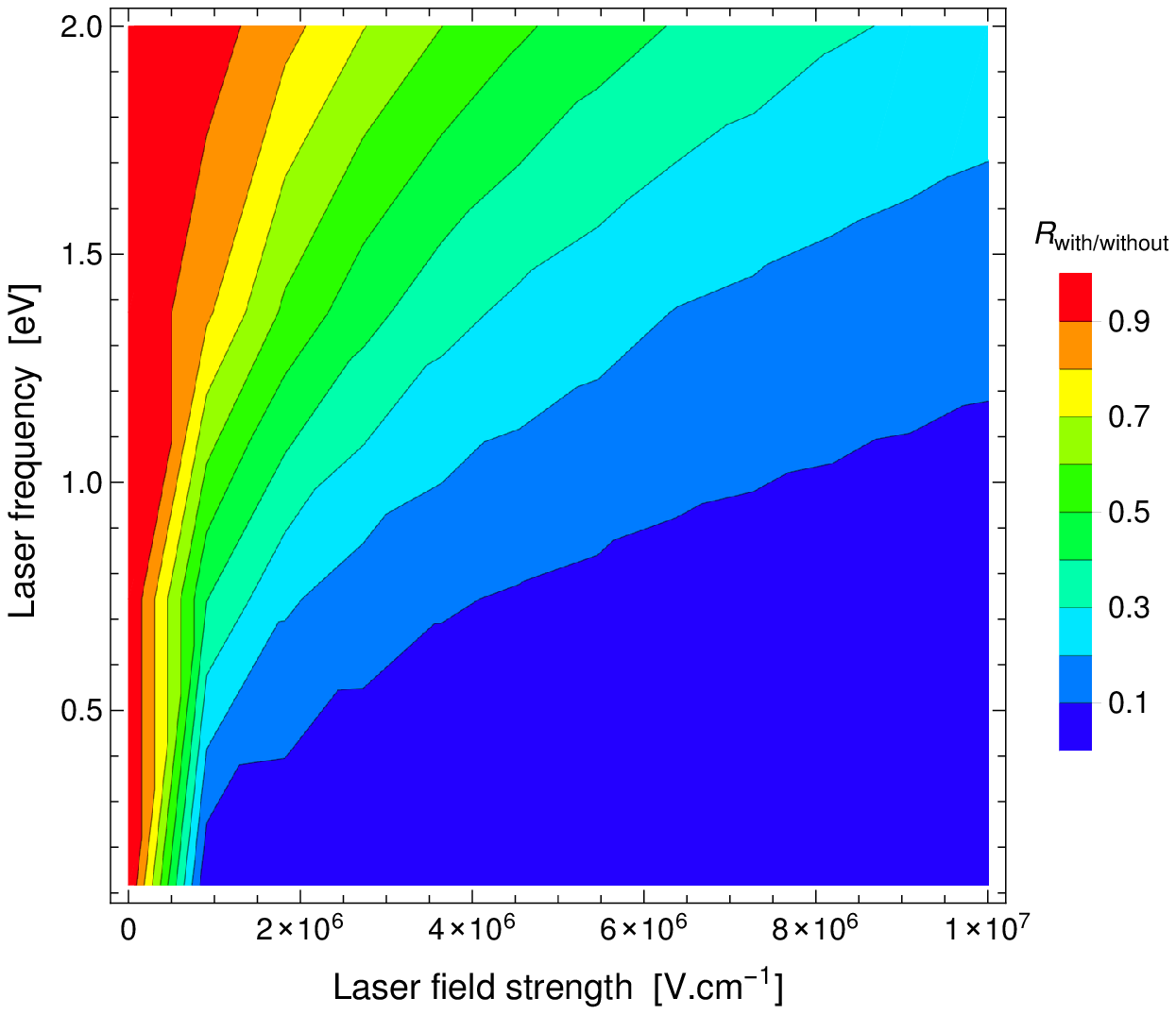}
    \caption{The 3-dimensional contour-plots of the same quantity as in figure \ref{fig8} as a function of the laser field strength $\mathcal{E}_{0}$ and frequency $\omega$ for $-20\leq s\leq+20$ exchanged photons.}
    \label{fig12}
  \end{minipage} %
  \hspace*{0.5cm}
  \begin{minipage}[t]{0.485\textwidth}
  \centering
    \includegraphics[width=\textwidth]{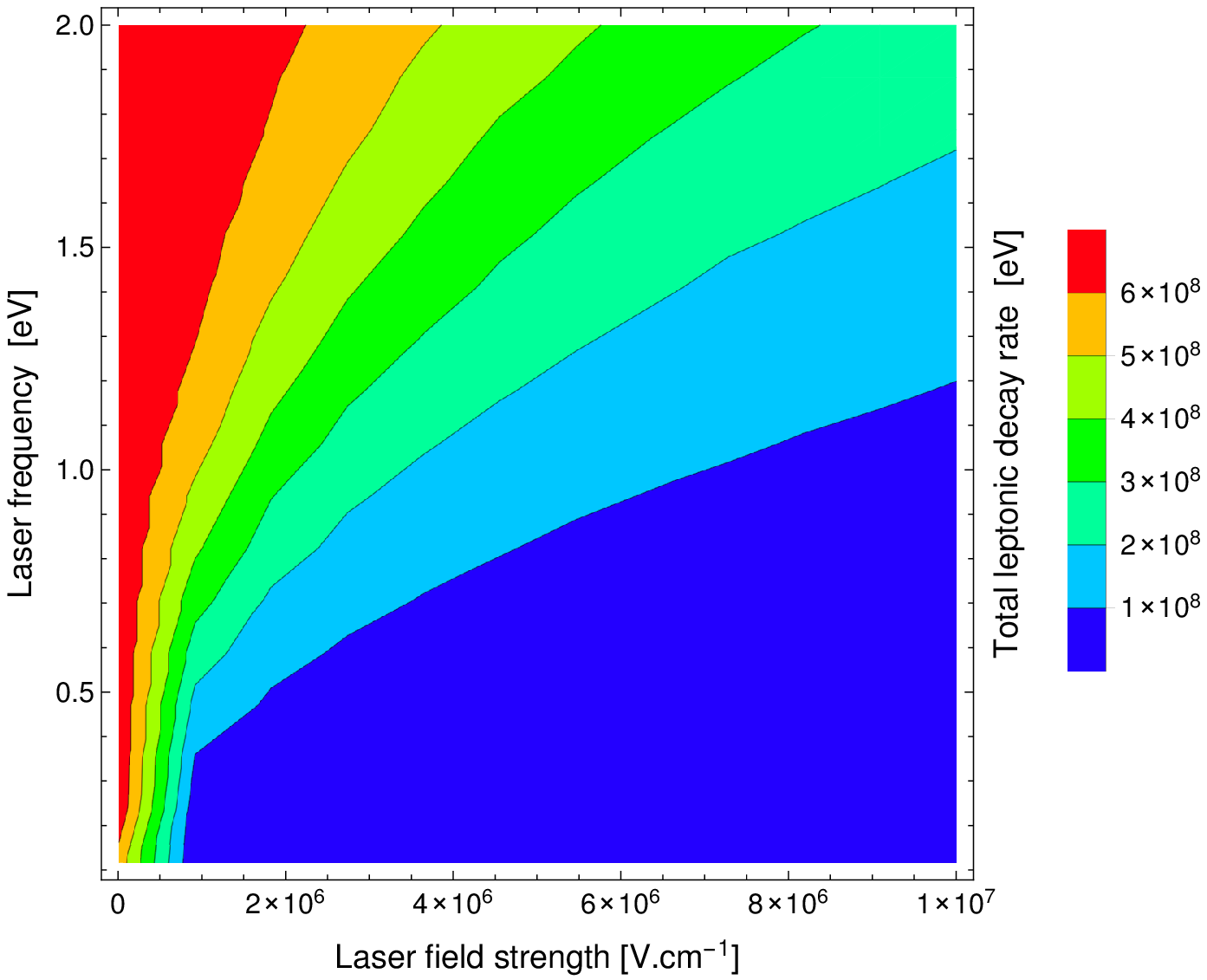}
    \caption{The 3-dimensional contour-plots of the same quantity as in figure \ref{fig9} as a function of the laser field strength $\mathcal{E}_{0}$ and frequency $\omega$ for $-20\leq s\leq+20$ exchanged photons.}
    \label{fig13}
  \end{minipage}
\end{figure}
%\begin{figure}[hbtp]
%\centering
%  \subcaptionbox{\label{a6}}{\includegraphics[height=5cm,width=.47\linewidth]{fig12}}\hspace*{0.5cm}
%  \subcaptionbox{\label{b6}}{\includegraphics[height=5cm,width=.47\linewidth]{fig13}}
%  \caption{The 3-dimensional contour-plots of the same quantities as in figure \ref{fig4} as a function of the laser field strength $\mathcal{E}_{0}$ and frequency $\omega$ for $-20\leq s\leq+20$ exchanged photons.}\label{fig6}
%\end{figure}
For illustration purposes, similarly to the 2-dimensional plot, the contour-plots in figures \ref{fig12} and \ref{fig13} show more information on the overall variation and shape of the ratio $R_{with/without}$ and total leptonic decay rate with respect to both laser field strength and frequency for an exchange of $-20\leq s\leq+20$ photons. This type of graphics is only a kind of 3-dimensional representation, in which one studies a certain quantity and its behavior by changing simultaneously two variables on a 2-dimensional plane and using colors to represent the variations of the quantity. Apparently, these two figures \ref{fig12} and \ref{fig13} are divided into many colored regions separated by contours, where each contour separating two different regions corresponds to a value on the bar legend. Looking at these two figures, we notice that they have the same feature to some extent, which indicates that the two quantities represented there change in the same way in terms of field strength and frequency. For the variation with respect to the laser frequency, we observe that the effect of the laser field on the two quantities decreases at high frequencies, while it becomes significant with increasing laser field strength at each specific frequency. This is completely in agreement with everything we said above. The important question now is how can we explain the great effect that the EM field has at strong field strengths. From figure \ref{fig10}, it appears that the value reached by the ratio $R_{with/without}$, when the number of exchanged photons is $-100\leq s\leq+100$ and at $\mathcal{E}_{0}=10^{8}~\text{V/cm}$, is $R_{with/without}=0.07$. This means that, the laser-assisted leptonic decay rate is $93\%$ lower than its corresponding one in the absence of the laser field. This is a significant effect of the laser on the decay rate. A computer calculation shows, when summing over $-10\leq s\leq+10$, that the minimum value reached by the quantity $R_{with/without}$ is $R_{with/without}^{min}=2.11603\times 10^{-10}$ at field strength $\mathcal{E}_{0}=10^{16}~\text{V/cm}$ (Schwinger limit), meaning that the quantity $R_{with/without}$ is almost zero at this field strength. This means that the decay rate in the presence of the laser is negligible compared to its equivalent one in the absence of the laser. Since the decay rate reflects, in its origin, the probability of occurrence of the decay, we conclude that the presence of the laser has considerably reduced the probability of occurrence of the leptonic $W^{-}$-boson decay if we do not exaggerate and say that it can suppress or prevent its decay. To our knowledge, there are two explanations for this behavior of the $W$ boson in the presence of the laser. The first is that the laser has suppressed these leptonic channels to then completely allow other hadronic ones, whether they are allowed in the absence of laser, such as $W^{-}\rightarrow \bar{u}d,~\bar{c}s$, or not, such as $W^{-}\rightarrow \bar{t}b$, especially since the $W$ boson gains in the EM field an excess of mass (effective mass) which increases with increasing field strength (e.g. $M_{W}^{*}=186.825~\text{GeV}$ at $\mathcal{E}_{0}=10^{16}~\text{V/cm}$ and $\hbar\omega=1.17~\text{eV}$). This first explanation can only be confirmed by studying the $W^{-}$ hadronic decay in the presence of the EM field. The second explanation is to attribute this behavior to the so-called quantum Zeno effect. The quantum Zeno effect can generally be defined by the inhibition of transitions between quantum states by repeated measurements of the state. This effect has attracted a lot of attention over the past years \cite{firstzeno,Hindered,zeno1,zeno2,zeno3,zeno4}. In $1977$, Misra and Sudarshan showed, based on quantum measurement theory, that an unstable particle would never decay when observed continuously \cite{firstzeno}. They were the first to give the name to this effect. In our case, the process of the $W$ leptonic decay is observed by inserting it into an EM field. Thus the laser as an external field here plays the role of the measurement instrument. In addition, the lifetime of the $W$ boson will certainly, as a result, be extended and long \cite{Hindered}. In this context, we want to link our results and compare them with the results obtained by Kurilin in $2004$ when he calculated the partial width of the $W$-boson leptonic decay in a strong EM crossed field \cite{kurilin2004}. He studies the changes of the decay width in terms of a characteristic parameter $\varkappa=eM_{W}^{-3}\sqrt{-(F_{\mu\nu}q^{\nu})^{2}}$, where $F_{\mu\nu}$ is the EM tensor. He has revealed that the partial decay width is a non-monotonic function of the field-strength parameter $\varkappa$. In superstrong fields $(\varkappa\gg 1)$, the partial width is greater than the corresponding one in a vacuum by a factor of a few tens, which means that the laser-assisted leptonic decay occurs faster than that in the absence of a laser. This result he reached does not constitute any contradiction with what we obtained. In contrast to the Zeno effect, there is an opposite effect to it called the quantum anti-Zeno effect, where the decay of some quantum systems from an initial state might be accelerated by frequent interrogations when the system remains in that state \cite{antizeno1,antizeno2,antizeno3}. This effect was discovered by Kofman and Kurizki \cite{kofman1,kofman2}. However, these two results, despite the difference between them due to the different approach followed in each of them, can be widely accepted with reference to the quantum Zeno effect or the anti-Zeno effect, depending on whether the decay is suppressed or enhanced.
\section{Conclusion}\label{sec:conclusion}
A detailed theoretical calculation of the leptonic $W^{-}$-boson decay $(W^{-}\rightarrow \ell^{-}\bar{\nu}_{\ell})$ in the presence of a circularly polarized electromagnetic field has been performed and an exact expression for the laser-assisted leptonic decay rate has been analytically derived. Our obtained numerical results show that the probability of decay to occur in the presence of a strong electromagnetic field is very small and almost non-existent, as if the laser at strong strengths and appropriate frequency is trying to stop the boson and delay its decay. This behavior can be considered as a result of the so-called quantum Zeno effect. As a consequence, the lifetime of the $W^{-}$-boson will also be affected by the laser and it will be long and extended, and it is not excluded that the branching ratios are also altered by the presence of the laser. In any case, this issue appears to be of particular interest and deserves a dedicated investigation which we will undertake in a forthcoming work to study the hadronic decay of the  $W^{-}$-boson.


\begin{thebibliography}{99}
\bibitem{salamin} Salamin Y I, Hu S X, Hatsagortsyan K Z and Keitel C H 2006 Relativistic high-power laser-matter interactions \emph{Phys. Rep.} \textbf{427} 41
\bibitem{bahk} Bahk S -W \textit{et al} 2004 The generation and characterization of the highest laser intensity $(10^{22}\text{W/cm}^{2})$ \emph{Opt. Lett.} \textbf{29} 2837
\bibitem{yanovski} Yanovsky V \textit{et al} 2008 Ultra-high intensity-$300$-TW laser at $0.1$ Hz repetition rate \emph{Opt. Express} \textbf{16} 2109
\bibitem{baifei} Baifei Shen and Yu M Y 2002 High-intensity laser-field amplification between two foils \emph{Phys. Rev. Lett.} \textbf{89} 275004
\bibitem{tajima} Tajima T and Mourou G 2002 Zettawatt-exawatt lasers and their applications in ultrastrong-field physics \emph{Phys. Rev. ST Accel. Beams} \textbf{5} 031301
\bibitem{bulanov} Bulanov S V, Esirkepov T and Tajima T 2003 Light intensification towards the Schwinger limit \emph{Phys. Rev. Lett.} \textbf{91} 085001
\bibitem{pukhov} Gordienko S, Pukhov A, Shorokhov O and Baeva T 2005 Coherent focusing of high harmonics: a new way towards the extreme intensities \emph{Phys. Rev. Lett.} \textbf{94} 103903
\bibitem{qed1} Ehlotzky F, Krajewska K and Kamiński J Z 2009 Fundamental processes of quantum electrodynamics in laser fields of relativistic power \emph{Rep. Prog. Phys.} \textbf{72} 046401
\bibitem{qed2} Di Piazza A,  M\"{u}ller C, Hatsagortsyan K Z and Keitel C H 2012 Extremely high-intensity laser interactions with fundamental quantum systems \textit{Rev. Mod. Phys.} \textbf{84} 1177
\bibitem{hartin} Hartin A 2018 Strong field QED in lepton colliders and electron/laser interactions \emph{Int. J. Mod. Phys. A} \textbf{33} 1830011
\bibitem{ritus1} Ritus V I 1985 Quantum effects of the interaction of elementary particles with an intense
electromagnetic field \emph{J. Russ. Laser Res.} \textbf{6} 497
\bibitem{ritus2} Nikishov A I and Ritus V I 1964 Quantum processes in the field of a plane electromagnetic wave and in a constant field 1 \emph{Sov. Phys. JETP} \textbf{19} 529
\bibitem{ehlotzky} Ehlotzky F 1985 Scattering phenomena in strong radiation fields II \emph{Can. J. Phys.} \textbf{63} 907
\bibitem{ritus3} Ritus V I 1969 Effect of an electromagnetic field on decays of elementary particles \emph{Zh. Eksp. Teor. Fiz.} \textbf{56} 986
\bibitem{becker} Becker W \textit{et al} 1983 A note on total cross sections and decay rates in the presence of a laser field \emph{Phys. Lett. A} \textbf{94} 131
\bibitem{ua1w} UA1 Collaboration 1983 Experimental observation of isolated large transverse energy electrons with associated missing energy at $s=540~\text{GeV}$ \emph{Phys. Lett.} \textbf{122B} 103
\bibitem{ua2w} UA2 Collaboration 1983 Observation of single isolated electrons of high transverse momentum in events with missing transverse energy at the CERN $\bar{p}p$ collider \emph{Phys. Lett.} \textbf{122B} 476
\bibitem{ua1z} UA1 Collaboration 1983 Experimental observation of lepton pairs of invariant mass around $95~\text{GeV}/c^{2}$ at the CERN SPS collider \emph{Phys. Lett.} \textbf{126B} 398
\bibitem{ua2z} UA2 Collaboration 1983 Evidence for $Z^{0}\rightarrow e^{+}e^{-}$ at the CERN $\bar{p}p$ collider \emph{Phys. Lett.} \textbf{129B} 130
\bibitem{glashow} Glashow S L 1980 Towards a unified theory: Threads in a tapestry \emph{Rev. Mod. Phys.} \textbf{52} 539
\bibitem{weinberg} Weinberg S 1980 Conceptual foundations of the unified theory of weak and electromagnetic interactions \emph{Rev. Mod. Phys.} \textbf{52} 515
\bibitem{salam} Salam A 1980 Gauge unification of fundamental forces \emph{Rev. Mod. Phys.} \textbf{52} 525
\bibitem{rubbia} Rubbia C 1985 Experimental observation of the intermediate vector bosons $W^{+}$, $W^{-}$, and $Z^{0}$ \emph{Rev. Mod. Phys.} \textbf{57} 699
\bibitem{vander} Van der Meer S 1985 Stochastic cooling and the accumulation of antiprotons \emph{Rev. Mod. Phys.} \textbf{57} 689
\bibitem{pdg2020} Zyla P A \textit{et al} (Particle Data Group) 2020 Review of Particle Physics \emph{Prog. Theor. Exp. Phys.} \textbf{2020} 083C01
\bibitem{kurilin2004} Kurilin A V 2004 Leptonic decays of the $W$ boson in a strong electromagnetic field \emph{Phys. Atom. Nucl.} \textbf{67} 2095
\bibitem{abramowitz} Abramowitz M and Stegun I A 1968 \emph{Handbook of Mathematical Functions} (Dover, New York)
\bibitem{obukhov1} Obukhov I A, Perez-Fernandez V K and Khalilov V R 1987 Electroweak lepton decay in the external field of a planar electromagnetic wave with circular polarization \emph{Russ. Phys. J.} \textbf{30} 383
\bibitem{kurilin1999} Kurilin A V 1999 Particle physics in intense electromagnetic fields \emph{Nuovo Cimento Soc. Ital. Fiz.} \textbf{112D} 977
\bibitem{obukhov2} Obukhov I A, Perez-Fernadez V K and Khalilov V 1984 Vector field equations in external electromagnetic fields \emph{Russ. Phys. J.} \textbf{26} 1117
\bibitem{bsm1} Barnett R M, Lackner K S and Haber H E 1983 Discovering supersymmetric particles in $W$-boson decay and $e^{+}e^{-}$ annihilation \emph{Phys. Rev. Lett.} \textbf{51} 176
\bibitem{bsm2} Barnett R M and Haber H E 1985 Detection of supersymmetric particles in $W$-boson decay \emph{Phys. Rev. D} \textbf{31} 85
\bibitem{mouslih} Mouslih S, Jakha M, Taj S, Manaut B and Siher E 2020 Laser-assisted pion decay \emph{Phys. Rev. D} \textbf{102} 073006
\bibitem{jakha} Jakha M, Mouslih S, Taj S and Manaut B 2021 Laser effect on the final products of $Z$-boson decay \emph{Laser Phys. Lett.} \textbf{18} 016002
\bibitem{greiner} Greiner W and M\"{u}ller B 2000 \textit{Gauge Theory of Weak Interactions} (Berlin: Springer)
\bibitem{volkov} Volkov D M 1935 On a class of solutions of the Dirac equation \emph{Z. Phys.} \textbf{94} 250
\bibitem{landau} Berestetskii V B, Lifshitz E M and Pitaevskii L P 1982 \emph{Quantum Electrodynamics} (Oxford U.K.: Butterworth-Heinemann) 
\bibitem{feyncalc1} Mertig R, B\"{o}hm M and Denner A 1991 Feyn Calc - Computer-algebraic calculation of Feynman amplitudes \emph{Comput. Phys. Commun.} \textbf{64} 345
\bibitem{feyncalc2} Shtabovenko V, Mertig R and Orellana F 2016 New developments in FeynCalc 9.0  \emph{Comput. Phys. Commun.} \textbf{207} 432
\bibitem{feyncalc3} Shtabovenko V, Mertig R and Orellana F 2020 FeynCalc 9.3: New features and improvements \emph{Comput. Phys. Commun.} \textbf{256} 107478
\bibitem{szymanowski} Szymanowski C, Véniard V, Ta\"{i}eb R, Maquet A and Keitel C H 1997 Mott scattering in strong laser fields \emph{Phys. Rev. A} \textbf{56} 3846
\bibitem{liberakdar} Li S M, Berakdar J, Chen J and Zhou Z F 2003 Mott scattering in the presence of a linearly polarized laser field \emph{Phys. Rev. A} \textbf{67} 063409
\bibitem{sumrule} Kroll N M and Watson K M 1973 Charged-particle scattering in the presence of a strong electromagnetic wave \emph{Phys. Rev. A} \textbf{8} 804
\bibitem{firstzeno} Misra B and Sudarshan E C G 1977 The Zeno's paradox in quantum theory \emph{J. Math. Phys.} \textbf{18} 756
\bibitem{Hindered} Mihokova E, Pascazio S and Schulman L S 1997 Hindered decay: Quantum Zeno effect through electromagnetic field domination \emph{Phys. Rev. A} \textbf{56} 25
\bibitem{zeno1} Koshino K and Shimizu A 2005 Quantum Zeno effect by general measurements \emph{Phys. Rep.} \textbf{412} 191
\bibitem{zeno2} Itano W M, Heinzen D J, Bollinger J J and Wineland D J  1990 Quantum Zeno effect \emph{Phys. Rev. A} \textbf{41} 2295
\bibitem{zeno3} Petrosky T, Tasaki S and Prigogine I 1991 Quantum Zeno effect \emph{Physica A}, \textbf{170} 306
\bibitem{zeno4} Panov A D 2001 Quantum Zeno effect in spontaneous decay with distant detector \emph{Phys. Lett. A} \textbf{281} 9  
\bibitem{antizeno1} Gontys V and Kaulakys B 1997 Quantum anti-Zeno effect \emph{Phys. Rev. A} \textbf{56} 1131
\bibitem{antizeno2} Lewenstein M and Rz\c a\.zewski K 2000 Quantum anti-Zeno effect \emph{Phys. Rev. A} \textbf{61} 022105
\bibitem{antizeno3} Fischer M C, Gutiérrez-Medina B and Raizen M G 2001 Observation of the quantum Zeno and anti-Zeno effects in an unstable system \emph{Phys. Rev. Lett.} \textbf{87} 040402
\bibitem{kofman1} Kofman A G and Kurizki G 1996 Quantum Zeno effect on atomic excitation decay in resonators \emph{Phys. Rev. A} \textbf{54} R3750
\bibitem{kofman2} Kofman A G and Kurizki G 2000 Acceleration of quantum decay processes by frequent observations \emph{Nature} \textbf{405} 546

\end{thebibliography}
\end{document}